\documentclass[11pt,hyper,letter]{JHEP3}
\usepackage{graphicx,amssymb,amsmath,amsfonts,cite}

\usepackage[english]{babel}
\usepackage{amssymb}
\usepackage{amsfonts}
\usepackage{amsmath}
\usepackage{graphicx}
\usepackage{epsfig}
\usepackage{cite}
\newcommand{\cE}{{\cal E}}

  \newcommand{\cL}{{\cal L}}

\newcommand{\tF}{\tilde{F}}
\newcommand{\tE}{{\tilde{E}}}

\newcommand{\be}{\begin{equation}} \newcommand{\ee}{\end{equation}}
\newcommand{\bea}{\begin{eqnarray}} \newcommand{\eea}{\end{eqnarray}}
\newcommand{\beann}{\begin{eqnarray*}}  \newcommand{\eeann}{\end{eqnarray*}}
\newcommand{\bfig}{\begin{figure}} \newcommand{\efig}{\end{figure}}
\newcommand{\ba}{\begin{array}} \newcommand{\ea}{\end{array}}
\newcommand{\bcen}{\begin{center}} \newcommand{\ecen}{\end{center}}
\newcommand{\btab}{\begin{tabular}} \newcommand{\etab}{\end{tabular}}

\def\tr{\operatorname{tr\:}}     \def\Tr{\operatorname{Tr\,}}

\def\a{\alpha}
\def\IR{\relax{\rm I\kern-.18em R}}
\newtheorem{Proposition}{Proposition}[section]

\newtheorem{Theorem}{Theorem}[section]
\newtheorem{Lemma}{Lemma}[section]

\newcommand{\bp}{\begin{Proposition}}   \newcommand{\ep}{\end{Proposition}}
\newcommand{\bt}{\begin{Theorem}}   \newcommand{\et}{\end{Theorem}}
\newcommand{\bl}{\begin{Lemma}}     \newcommand{\el}{\end{Lemma}}
\newcommand{\bc}{\begin{Corolary}} \newcommand{\ec}{\end{Corolary}}
\def\pa{\partial}
\def\CR{\nonumber\\}

\def\d{\partial}

\title{Deformation Constraints on Solitons and  D-branes}

\author{Sophia K. Domokos${}^1$\footnotemark[1], Carlos Hoyos${}^2$\footnotemark[2], Jacob Sonnenschein${}^2$\footnotemark[3]\,
\\
${}^1$\textit{Weizmann Institute of Science, Rehovot 76100, Israel}\\
\\
${}^2$ \textit{
  Raymond and Beverly Sackler Faculty of Exact Sciences \\
School of Physics and Astronomy \\
Tel-Aviv University, Ramat-Aviv 69978, Israel.}
}

\footnotetext[1]{E-mail address: \email{sophia.domokos@weizmann.ac.il}}
\footnotetext[2]{E-mail address: \email{choyos@post.tau.ac.il}}
\footnotetext[3]{E-mail address: \email{cobi@post.tau.ac.il}}

\abstract{We derive a set of constraints on soliton solutions using
geometric deformations, and transformations by internal symmetries
with space-dependent parameters. We show that Derrick's theorem and
a more complete set of constraints due to Manton are special cases
of these deformation constraints (DC). We demonstrate also that
known soliton solutions obey the DC, and extract novel results by
applying the constraints to systems of D-branes, taking into account
both Dirac-Born-Infeld and Wess-Zumino actions, and examining cases
with and without D-brane gauge fields. We also determine a relation
with the Hamiltonian constraint for gravitational systems, and
discuss configurations of finite extent, like Wilson lines. }

\preprint{TAUP-2968/13\\ WIS/08/13-JUN-DPPA}

\begin{document}

\section{Introduction}

Solitons are defined as  finite-energy static solutions to the
classical equations of motion. They play an important role in a wide
range of physical systems, from non-linear optics to particle
physics. Famous examples include the kink solution in the
sine-Gordon theory \cite{Coleman} and other integrable
two-dimensional models, vortices of the Abelian-Higgs model
\cite{Nielsen:1973cs,de Vega:1976mi}, magnetic monopoles
\cite{'tHooft:1974qc,Polyakov:1974ek}, skyrmions
\cite{Skyrme:1962vh,Adkins:1983ya} and many others.

In recent years, static solutions of D-branes, strings and
gravitational actions have drawn increased interest.  These
solutions often have infinite energy, but in certain cases there
soliton solutions do exist. One example is the dual description of a
baryon in the Sakai-Sugimoto model, which is realized as an
instanton on the D8 branes \cite{Hata:2007mb}. Both the `` old
solitons" and those analyzed more recently are solutions of
non-linear differential equations for which analytic solutions are
very scarce. Conditions and constraints on such solutions serve as
useful tools that guide one in selecting tractable ans\"{a}tze or
restricting the existence of solitonic states. The well-known
Derrick-Hobart theorem  \cite{Derrick:1964ww,Hobart} is a
prototypical example of such a constraint: it shows that scalar
field theories with two derivatives can have soliton solutions only
in one space dimension. More recently, Manton \cite{Manton:2008ca}
developed an extended class of constraints for solitons. According
to these conditions, the space integral of the stress tensor
associated with static configurations must vanish for fields
decaying sufficiently fast near the boundary of the space.

 In \cite{Manton:2008ca} these conditions were applied to the Skyrme system, monopoles and vortices in Higgs models.
Various aspects of Derrick's theorem  were also discussed in
\cite{Jackiw:1976pq,Faddeev:2000qw,Harland:2010wc,Endlich:2010zj,Padilla:2010ir}.

In this work we derive the constraints of
\cite{Derrick:1964ww,Hobart} and of \cite{Manton:2008ca} from a new
perspective: infinitesimal geometrical deformations of solitons.
Since a soliton is a stable solution to the equations of motion, it
must lie at a minimum of the energy. If we deform the soliton by
some small transformation, the deformed configuration's energy must
therefore be larger than the energy of the original soliton. This
analysis yields conditions to be satisfied by any theory with
soliton solutions, and can be used to impose constraints on these
solutions.

First we study ``geometrical'' deformations around soliton
solutions. We consider deformations which depend linearly  on
coordinates, like (nonisotropic) dilatations and shear
transformations. We show that the former yield a generalization of
Derrick's theorem and the latter lead to Manton's integral
conditions.

A second class of deformations  involves internal symmetries of the
theory, which are broken by soliton solutions. In complete analogy
with the ``geometrical'' constraints, we show that for
energy-minimizing solitons, the volume integral of the space
components of the conserved current is constrained. For systems with
spontaneous breaking of the global symmetries, namely systems with
Nambu-Goldstone bosons, the integral equals a finite surface term;
otherwise, it must vanish. We refer to the constraints derived from
both the geometrical and the internal symmetry deformations as
``deformation constraints'' (DC).

The bulk of this paper deals with the application of the DC to a
wide class of models. We begin with a ``warm-up:'' generalizations
of Derrick's work to sigma models and higher derivative scalar
actions. We then survey soliton systems in various dimensions and
check that DC associated with conserved currents are fulfilled.
These include (trivially) the topological currents, currents
associated with global symmetries in the Skyrme model, and with
local symmetries in the Abelian Higgs model and  the  't
Hooft-Polyakov monopole \cite{'tHooft:1974qc,Polyakov:1974ek}.

 We then derive the DC for scaling  deformations of the
Dirac-Born-Infeld (DBI) action of D-branes, including both scalars
and gauge fields on the brane: we write down the stress tensor of a
general embedding, with electric and magnetic fields. We also
introduce the Wess-Zumino (WZ) action, and compute the contributions to the
stress tensor for the D1, D2 and D3 brane models.
 We then explore some specific examples of these systems.

 Finally, we study properties of holographic MQCD\cite{Aharony:2010mi}. For flavor M5-branes in an
 M-theory background we find that Derrick's condition matches with the
vanishing
 of the momentum density associated to a spatial direction.
 This relation is then realized in a general setup with scalar fields.
 Finally we show that
 for gravitational backgrounds this coincides with the ``Hamiltonian constraint'' derived
 by
 using the ADM formalism along the spatial direction.

The paper is organized as follows: in section \S 2 we describe
deformations of static field configurations and deduce the DC. We
consider first geometrical deformations (shears and anisotropic
dilations), then deformations associated with internal symmetries.
In section \S 3 we describe several applications of the DC to
theories with soliton solutions. In \S 4 we apply these DC to
D-branes and strings, including both the DBI and WZ actions. In
section \S 5 we present some applications in string theory including
flavor branes in M-theory, gravitational backgrounds, and Wilson
lines.
 We summarize, conclude, and
raise open questions in section \S 6. Two appendices contain details
of known examples and small extensions:  probe branes in brane
backgrounds,
 D3-branes with electric and magnetic fields, and
the DC  in DBI theories  with scalars.

\section{Deformation constraints}\label{sec:defcon}

Solitons need not be the lowest energy states available,
 but they must lie at minima of the energy -- otherwise they can become locally unstable.
 From this simple statement one can derive a series
 of non-trivial constraints by slightly deforming the soliton solutions. In this work, we will study
 deformations of solitons under infinitesimal shears, dilatations, and internal symmetry transformations.

\subsection{Geometrical deformations of solitons}

Consider the simple case of a theory with one or more scalars
$\phi^a$, that possesses a  \textit{finite energy} static solution,
$\phi^a_0(x)$. How do small deformations affect the energy of this
solution?

Generically, the energy of the soliton is a function of the fields
and its derivatives,
\begin{equation}
E[\phi^a_0]=\int d^dx\,\cE(\phi^a_0,\partial_i\phi^a_0)
\end{equation}
where $\cE$ is the energy density. Let's say we deform the
configuration by a geometric deformation in space, $\Lambda x$. The
deform solution will look like
\begin{equation}
\phi^a_\Lambda(x)=\phi^a_0(\Lambda x).
\end{equation}
For small deformations we can expand
\begin{equation}
(\Lambda x)^i\simeq x^i+\xi^i(x).
\end{equation}
In what follows, we will consider only linear transformations of the
form
\begin{equation}
(\Lambda x)^i=\Lambda^i_{\ j}x^k +b^i\simeq x^i+\lambda^i_{\ k}x^k+b^i.
\end{equation}
These include rigid translations, $b^i$, rigid rotations when
$\lambda^i_{\ j}$ is antisymmetric, anisotropic dilatations when
$\lambda^i_{\ j}$ is diagonal, and volume-preserving shear
deformations when $\lambda^i_{\ j}$ is symmetric and has no diagonal
components.\footnote{Some of the anisotropic deformations are also
volume-preserving, the condition is that $\lambda^i_{\ j}$ is
traceless. If $\lambda^i_{\ j}$ is symmetric it can always be
diagonalized locally, so shear deformations are a combination of
rotations with anisotropic dilatations.} The underlying
translational and rotational symmetry of the theory implies that the
energy will not change under rigid translations and rotations, but
it can change otherwise.

The energy of the configuration that has suffered a small deformation is, to leading order,
\begin{align}
\notag E[\phi^a_\Lambda] &=\int d^dx\,\cE\left(\phi^a_\Lambda(x),\partial_i\phi^a_\Lambda(x)\right)\\
\notag &=\int d^d x' \left|\left|\frac{\delta x^i}{\delta {x^j}'}\right|\right|\cE\left(\phi^a_0(x'),\frac{\partial {x^j}'}{\partial x^i}\partial_j' \phi^a_0(x')\right)\\
\notag &\simeq \int d^d x'\cE\left(\phi_0^a(x'),\partial_i \phi^a_0(x')\right)+\int d^d x'\, \partial_i\xi^j\left[\delta^i_{\ j}\cE-\frac{\delta \cE}{\delta \partial_i \phi^a}\partial_j\phi^a_0\right]\\
&=E[\phi^a_0]-\int d^d x'\, \partial_i\xi^j \Pi^i_{\ j}(\phi^a_0).
\end{align}
In the second line we have made the change of variables ${x^i}'=(\Lambda x)^i$ and in the next lines we have expanded for $x^i\simeq {x^i}'-\xi^i$.
Here $\Pi^i_{\ j}$ is a stress tensor for static configurations, based on the energy density $\cE$,
\begin{align}
\Pi^i_{\ j}=\frac{\delta \cE}{\delta \d_i\phi^a}\d_j\phi^a-\delta^i_j\cE~.
\end{align}
The difference in the energy of the deformed solution compared to
the original one is given by the stress tensor ($\Pi^i_{\ j}$) evaluated
at the soliton solution:
\begin{equation}
E[\phi^a_\Lambda]-E[\phi^a_0]=\int d^d x \ \delta\cE=-\int d^d x\,
\partial_i\xi^j \Pi^i_{\ j}.
\end{equation}
The variation of energy for a transformation linear in the
coordinates becomes
\begin{equation}\label{eq:Ediff}
E[\phi^a_\Lambda]-E[\phi^a_0]=- \lambda^j_{\ i}\int d^d x\, \Pi^i_{\
j},
\end{equation}
If the integral on the right-hand-side (RHS) of \eqref{eq:Ediff} is
non-vanishing, we can always choose $\lambda^i_{\ j}$ in such a way
that the RHS becomes negative. This would mean that the transformed
configuration is energetically favored over the original!  For
stable configurations, therefore, the integral should vanish:
\begin{equation}\label{manton}
\int d^d x\, \Pi^i_{\ j}=0.
\end{equation}
This is automatically satisfied for static solutions to the
equations of motion, up to boundary terms. The stability of static
solutions would then be determined by the second order term, which
should be positive if the soliton is a minimum energy solution.

The conditions \eqref{manton} were first derived (in a different
way) by Manton \cite{Manton:2008ca}, where they were used to study
monopoles, instantons, and Skyrmions. A linear combination of these
conditions  gives the vanishing of the trace of the stress tensor
\begin{equation}\label{derrick}
\int d^d x\, \Pi^i_{\ i}=0.
\end{equation}
This corresponds to an isotropic dilatation, the deformation used by
Derrick to prove that no soliton solutions exist for $d>1$ in scalar
theories with positive potentials \cite{Derrick:1964ww,Hobart}. The
proof, in essence, shows that $\Pi^i_{\ i}\geq 0$, so $\int \Pi^i_{\
i}=0$ only for the trivial solution. Physically \eqref{derrick} can
be interpreted as the virial theorem for field theories (see e.g.
\cite{Faddeev:2000qw,Gibbons:1997xz}). One can use \eqref{manton} to
impose more restrictive conditions on the field theory action or on
the properties of the soliton solutions, as algebraic relations
between derivatives of the fields and the potential.

It is important to note that in theories with only scalar fields,
the energy density functional of a static configuration coincides
with minus the Lagrangian density $\cE=-\cL$, and $\Pi^i_{\ j}$
coincides with minus the spatial part of the energy-momentum tensor
\begin{align}
\Pi^i_{\ j}=-T^i_{\ j} = -\frac{\delta \cL}{\delta \d_i\phi^a}\d_j\phi^a+\delta^i_j\cL~
\end{align}
This is not always the case, however. For instance, in gauge
theories with  electric fields turned on, the energy is a Legendre
transform of the Lagrangian. Furthermore,  the energy density
functional is not unequivocally defined. It is identified with the
time component of the energy-momentum tensor $\cE=T^0_{\ 0}$, which
can be modified by adding improvement terms. These terms do not
affect the conservation equations but alter some properties of the
energy-momentum tensor. For instance, in Maxwell's theory the
canonical energy-momentum tensor is not gauge invariant or positive
definite: this can be fixed adding an improvement term. Generically,
\begin{equation}
\cE=T^0_{{\rm can}\, 0}+\partial_i \Psi^{i 0}_{\ \ 0},
\end{equation}
where the improvement term $\Psi$ satisfies
\begin{equation}
\Psi^{\rho \mu}_{\ \ \nu}=-\Psi^{\mu\rho}_{\ \ \nu}.
\end{equation}
This guarantees conservation of the (improved) energy-momentum
tensor. For static configurations (that is, where the time
derivatives of all fields vanish)
\begin{equation}
\cE=-\cL+\partial_i \Psi^{i 0}_{\ \ 0}.
\end{equation}
Using the fact that $T^i_{ \ j}=T^i_{{\rm can}\, j}+\partial_k
\Psi^{ki}_{\ \ j}$ for static solutions, we find
\begin{equation}
\delta\cE=-\partial_i\xi^j \Pi^i_{\ j}=\partial_i\xi^j T^i_{\ j}-\partial_i\left( \delta\Psi^{i 0}_{\ \ 0}+\partial_k \Psi^{ki}_{\ \ j}\xi^j\right).
\end{equation}
Choosing an appropriate energy density functional will be an
important part of the analysis. We will keep the notation $\Pi^i_{\
j}$ to denote the stress tensor derived from the energy functional,
to avoid confusion with the energy-momentum tensor (even in the
cases where they are interchangeable).

The conditions \eqref{manton} rely crucially on the fields dying off
quickly at infinity. For a static solution, the conservation of the
energy-momentum tensor implies that $\partial_i T^i_{\ j}=0$. It is
then possible to express the variation as a surface term:
\begin{align}
\notag \int d^d x\, \partial_i\xi^j \Pi^i_{\ j} &=\int d^d
x\partial_i\left[- T^i_{\ j}\xi^j+\delta\Psi^{i 0}_{\ \
0}+\partial_k \Psi^{ki}_{\ \ j}\xi^j\right]\\ \label{tijsurface}
&=\oint d^{d-1}\sigma_i\,\left[ -T^i_{\ j}\xi^j+\delta\Psi^{i 0}_{\
\ 0}+\partial_k \Psi^{ki}_{\ \ j}\xi^j\right],
\end{align}
which does not necessarily vanish for configurations that extremize
the energy, because the transformations we use do not necessarily
vanish at infinity. The transformations in such cases can thus
affect the boundary conditions of the fields, if they fields so not
die off sufficiently fast. Since it is the action \textit{together
with} the boundary conditions that defines the physical system,
transformations that alter the boundary conditions relate two
different physical systems, rather than comparing different field
configurations within the same system. We will explain this issue in
more detail when we discuss deformations of solitons by global
symmetries in the next subsection.

\subsection{BPS conditions and the stress tensor}

Bogomolnyi-Prasad-Sommerfield (BPS) conditions
\cite{Bogomolny:1975de,Prasad:1975kr} are lower bounds on the energy
of solutions to the equations of motion, as a function of their
topological charge. They are very useful for finding soliton
solutions, solutions that saturate the BPS bound can by found by
solving a much simpler set of first order equations (rather than the
second order equations of motion).

When a soliton saturates a BPS condition, not only the integral of
the stress tensor \eqref{manton}, but the components of the stress
tensor themselves vanish. It was argued in \cite{Moreno:2008me} that
this can be understood as a consequence of supersymmetry, even in
purely bosonic models. Consider for instance a scalar field  $\phi$
in $1+1$ dimensions with a positive semi-definite potential. The
supersymmetric extension of the theory is
\begin{equation}
S=\int d^2 x\left( \frac{1}{2}\partial_\mu \phi\partial^\mu\phi+\frac{i}{2}\bar{\psi}\gamma^\mu\partial_\mu \psi+\frac{1}{2}F^2+F W(\phi)-\frac{1}{2}W'(\phi)\bar{\psi}\psi\right)\,,
\end{equation}
with $\psi$ a Majorana spinor and $F$ an auxiliary field.  The spatial components of the stress tensor for a static bosonic configuration are
\begin{equation}
T_{01}=T_{10}=0, \ \
T_{11}=\frac{1}{2}(\partial_1\phi)^2-\frac{1}{2}W^2.
\end{equation}
This coincides with the ordinary bosonic theory for the potential
$V=W^2/2$.

Supersymmetry relates the components of the stress tensor to the components of the supercurrent
\begin{equation}
S_\alpha^\mu=(\gamma^\nu\partial_\nu\phi+i W)\gamma^\mu \psi,
\end{equation}
through the supersymmetric Ward identity
\begin{equation}
\left\{S_\alpha^\mu, \,\bar{Q}_\beta \right\}=2i\gamma^\nu T_\nu^{\ \mu}+2i{\gamma_3}_{\alpha\beta} W'\epsilon^{\mu\nu}\partial_\nu\phi.
\end{equation}
The last term is proportional to the topological current (whose
time component integrates to the central charge). From this
expression, one can use the fact that the BPS solution is invariant
under half of the supercharges to show that the components of the
stress tensor vanish. This was checked explicitly in
\cite{Moreno:2008me} for $1+1$ dimensional scalars, and also for
Abelian vortices in $2+1$ dimensions, but the argument can in
principle be generalized to any supersymmetric theory with BPS
soliton solutions.

\subsection{Deformations by global symmetries}

One can generalize the conditions \eqref{manton}  for geometrical
deformations to deformation of global symmetries broken by the
solitonic configuration. Consider a global symmetry with associated
global conserved current $J^\mu$. For static solutions, current
conservation implies $\partial_i J^i=0$. Following the same logic as
before, we can deform the soliton solution $\phi^a(x)$ by a
transformation
\begin{equation}
\delta_\theta \phi^a(x)=\theta^A(x) {T_A}^a_b(\phi^b(x)),
\end{equation}
where ${T_A}$ are generators of the symmetry group in the
appropriate representation. The
variation of the energy is
\begin{equation}
\delta_\theta E[\phi^a]=\int d^d x\,\left[\partial_i \theta^A
J_A^i+\partial_i (\delta_\theta \Psi^{i0}_{\ \ 0})\right]=\int d^d
x\,\partial_i \theta^A J_A^i,
\end{equation}
where for the moment we assume that the boundary integral vanishes.
As before we can pick a transformation linear in the coordinates
$\theta^A=C^A_{\ i}x^i$, and for arbitrary $C^A_{\ i}$ we find the
conditions
\begin{equation}\label{currentcond}
\int d^d x\,J_A^i=0.
\end{equation}
This condition should be satisfied by solutions to the equations of
motion. For other configurations, there will be an instability such
that the charges tend to rotate along the spatial directions.

 It is easy to see, for instance, that \eqref{currentcond} is trivially satisfied for topological currents in field
theories, since the topological current  associated with solitons of
general dimension\cite{Frishman:2010zz} is proportional to the
Levi-Civita tensor. The space components of this current thus vanish
for static configurations and \eqref{currentcond} is simply zero, as
discussed at greater length in section \ref{app:global} below.

\subsection{Boundary Terms}

There is an important caveat that applies to both the conditions we
have just derived for  global internal symmetries and for
geometrical deformations via the energy-momentum tensor. The
variation of the energy has to vanish only up to a total derivative,
but the conservation of the energy-momentum tensor and the current
imply that the variations we study are total derivatives, as noted
in \eqref{tijsurface} and via a similar identity for the current:
\begin{align}\label{intsurface}
\notag &\int d^d x \partial_i \xi^j T^i_{\ j}=\oint d^{d-1}\sigma_i T^i_{\ j} \xi^j=\lim_{|x|\to \infty}\oint d^{d-1}x \hat{x}_i T^i_{\ j} \xi^j,\\
&\int d^d x \partial_i \theta^A J_A^i =\int d^d x \partial_i(
\theta^A J^i_A)=\oint d^{d-1}\sigma_i \theta^A J^i_A=\lim_{|x|\to
\infty}\oint d^{d-1}x \hat{x}_i \theta^A J^i_A.
\end{align}
There are possible boundary contributions from the improvement terms
as well. $\hat{x}^i$ is a unit vector along the $i$th spatial
direction, and the surface integral is taken over a sphere whose
radius goes to infinity. The integral over the energy-momentum
tensor of the global currents does not need to vanish: it could also
be a constant. This complication to \eqref{currentcond} occurs
because we have used transformations linear in the coordinates,
$\xi^i=\lambda^i_{ \ k}x^k$, $\theta^A=C^A_k x^k$, which introduce
factors that diverge as the volume of the space
\begin{equation}
\oint d^{d-1}x\,x^k \sim |x|^d.
\end{equation}
In order to have a finite integral, the conserved currents should
decay as
\begin{equation}
\hat{x}_iT^i_{\ j}\sim \frac{1}{|x|^d}, \ \ \hat{x}_i J_A^i\sim
\frac{1}{|x|^d}.
\end{equation}
For a global current, the conservation equation $\partial_i J_A^i=0$
implies that at leading order for large radii,
\begin{equation}
J_A^i=\frac{x^2\delta^{ij}-d x^i x^j}{|x|^{d+2}}v_{Aj}+\cdots,
\end{equation}
where $v_{Aj}$ is some constant. In fact, $J_A^i$ itself is a total
derivative
\begin{equation}
J_A^i=\partial^i\left(\frac{x^j v_{Aj}}{|x|^d}
\right)=\partial^i\phi_A,
\end{equation}
and solving the conservation equation is equivalent to solving the
equation for a massless scalar field $\partial^i
\partial_i\phi_A=0$ when the solutions are static. For the
energy-momentum tensor the situation is similar, except that the
scalar will be labeled by a spatial index, $v_{Aj}\to v_{kj}$,
$\phi_A\to \phi_k$.

The currents are proportional to the gradient of a scalar field when
the associated symmetry has been spontaneously broken,\footnote{That
is, not by the soliton itself but in the ground state. Properly speaking the symmetry is non-linearly realized in the action.} and there
are Goldstone bosons. From the argument above we see that the
conditions \eqref{manton} and \eqref{currentcond} do not apply to
those cases. The non-zero variation of the energy, in this case, is
a consequence of changing the boundary conditions of the field
configuration at infinity. For a Goldstone boson $\varphi^A$, a
global symmetry transformation shifts the field
\begin{equation}
\varphi^A\to \varphi^A+\theta^A.
\end{equation}
Since $\theta^A$ is linear in the coordinates, this modifies the
asymptotic behavior of the field, so the deformation maps the
original theory to a different physical system, not to a different
point in the configuration space of the same system. The requirement
of vanishing (or constant) integral for the stress tensor no longer
yields a valid stability check for the soliton.

We have a similar problem if there are boundaries at a finite
distance: linear deformations alter the boundary conditions, which
will be reflected in the boundary contribution to the variation of
the energy. One must then generalize the variational principle to
include the boundaries in the analysis. This is particularly
important for to the study open strings from the worldsheet
perspective, or branes with boundaries.

\section{Some applications in theories with soliton solutions}\label{Sociss}

In this section  we survey  field theories that admit soliton
solutions and analyze  space integrals over the spatial components
of the conserved currents. The latter include topological currents,
currents associated with global symmetries and with local
symmetries.

\subsection{Derrick's theorem}

Derrick's theorem is simply a consequence of \eqref{manton}, but we
will often make use of isotropic scaling deformations according to
Derrick's original framework. In many cases, this already allows one
to impose very strong constraints on possible soliton solutions, and
in some cases, exclude them entirely. We review this strategy now,
and point out some simple extensions to the original theorem. In
Appendix \ref{app:derrickbrane}, we also apply isotropic scaling to
the DBI action in some examples.

Consider the Lagrangian density of a scalar field in $d+1$
dimensions \be \label{Lagden} {\cal L}=- \frac12
\pa_\mu\phi\pa^\mu\phi - V(\phi), \ee where the potential is
non-negative and vanishes at its minima. (We use the mainly plus
metric convention.) The energy associated with this family of field
configurations is
\be
E= \int d^d x \left [ \frac12(\nabla\phi(\vec
x))^2 + V(\phi( \vec x))\right ].
\ee
Let us
consider a uniform scaling deformation of the soliton in this
system: $\phi(x^i)\rightarrow\phi(\lambda x^i)$ for some positive
real number, $\lambda$. The energy must be extremized on the stable
solution: $dE(\lambda=1)/d\lambda=0$. Note that for the energy to
actually be \textit{minimized} we also need $d^2E/d\lambda^2>0$.
The energy
associated with this  family of field configurations is \be
E(\lambda)= \int d^d x \left [ \frac12(\nabla\phi(\lambda \vec x))^2
+ V(\phi(\lambda \vec x))\right ]= \lambda^{-d}\int d^d x \left [
\frac12\lambda^2(\nabla\phi(\vec x))^2 + V(\phi(\vec x))\right ],
\ee where after performing the re-scaling of the soliton, we changed
variables in the integral of the energy, $x^\mu\rightarrow
x^\mu/\lambda$. The variation of the energy is thus \be
\label{derthe} \left.\frac{dE(\lambda)}{d\lambda}\right|_{\lambda=1} = -\int d^d
x \left [ \frac12 (d-2)(\nabla\phi(\vec x))^2 +d V(\phi( \vec
x))\right ]=0. \ee For $d>2$, each term must vanish separately. This
occurs only for the vacuum state. For $d=2$, it is  the potential
alone which must vanish, but again this occurs only for the vacuum.
Hence the statement of Derrick's theorem: {\it For $d\geq 2$ the
only non-singular time independent solution of finite energy is the
vacuum.}

One can quickly derive a similar result for sigma models with
Lagrangian \be {\cal L}= -\frac12 G^{ab}(\phi)\pa_\mu\phi_a\pa^\mu
\phi_b -V(\phi_a). \ee Repeating the procedure of scaling
$x\rightarrow \lambda x$ and demanding extremum for $\lambda =1$ one
finds \be \label{derthegen}
\left.\frac{dE(\lambda)}{d\lambda}\right|_{\lambda=1} = -\int d^d x \left [
\frac12 (d-2) G^{ab}\nabla\phi_a\nabla \phi_b +dV(\phi_a) \right ]=0~. \ee When the signature of the metric
$G^{ab}(\phi)$ is positive then the conclusions for the generalized
case are the same as for the flat space case; if the signature is
not positive, \eqref{derthegen} does not yield constraints in any
dimension $d\neq 2$.

\paragraph{Higher derivative actions}

Another interesting example is that of soliton solutions to differential
equations with more than two derivatives, such as the KdV and
non-linear Schr\"odinger equations. These  can
be derived as Euler-Lagrange equations of Lagrangians that include
terms of higher than the first derivative of the field.

Let us discuss some theories of this type, which appear (and have
been studied using Derrick's theorem) in the context of galileons
\cite{Endlich:2010zj,Padilla:2010ir}. First consider a Lagrangian
that includes second order derivatives of scalar field \be {\cal L}
= -\frac12[(\pa_\mu\phi)^2 + a (\pa_\mu\pa^\mu\phi)^2] -V(\phi)~.
\ee The corresponding equation of motion reads \be
\pa_\mu\pa^\mu\phi- a \pa^2 ( \pa^2\phi) +\frac{\pa V(\phi)}{\pa
\phi} =0~. \ee The action is invariant under space-time translations
$x^\mu\rightarrow x^\mu + a^\mu$ and the corresponding conserved
energy momentum tensor is given by \be T_{\mu\nu} = \pa_\mu\phi
\pa_\nu\phi + a[ \pa^2\phi\pa_\mu\pa_\nu \phi -
\pa_\mu(\pa^2\phi)\pa_\nu\phi ] - \eta_{\mu\nu} {\cal L}. \ee For
static configurations the Hamiltonian takes the form \be H= \int d^d
x \left [\frac12 ( \nabla \phi)^2 + \frac12 a ( \nabla^2 \phi)^2 +
V(\phi)\right ] . \ee Scaling $x\rightarrow \lambda x$ and requiring
extremality  for $\lambda=1$ we get \be
\left.\frac{dH}{d\lambda}\right|_{\lambda=1}= \int d^d x \left [\frac12 ( 2-d) (
\nabla \phi)^2 +   \frac12 a (4-d)  ( \nabla^2 \phi)^2 -d V(\phi)
\right ].
 \ee
 The higher derivative terms thus ease the restriction
on solitonic solutions for pure scalar field theories: we can get
solitons for $d<4$.

Generalizing this result to any higher order derivative Lagrangian
density, where the derivative terms are quadratic in the fields of
the form
\be
 {\cal L} = \frac12[(\pa_\mu\phi)^2 +\sum_n^N   a_n
(\pa^n\phi)^2] -V(\phi)],
\ee
one can see that the constraint in principle allows solitons
for any dimension $d<2N$.

\paragraph{Monopoles}

The conditions \eqref{manton} can also be used to verify the existence of soliton
solutions for systems involving gauge fields -- these checks are performed in \cite{Manton:2008ca} for
numerous examples; we present the analysis for monopoles.

To see how this works for gauge fields, consider a theory of a
scalar plus gauge fields. The Lagrangian density is \be {\cal L}=
-\frac14 F_{\mu\nu}F^{\mu\nu} - \frac12 (D_\mu\phi)^*(D^\mu\phi)-
V(\phi^*\phi) \ee where $D_\mu= \pa_\mu -ieA_\mu$ The corresponding
energy for static configurations ($F_{i0}=0$,
$D_0\phi=0$)  reads \be E= \int d^dx\left[ \frac14
F_{ij}F^{ij} + \frac12 (D_i\phi)^*(D^i\phi)+
V(\phi^*\phi)\right] \ee For $d=3$ the stress tensor is
\begin{equation}
T_{ij}=F_i^{\ \mu} F_{j\mu}+\frac{1}{2}\left(D_i\phi^*
D_j\phi+D_j\phi^* D_i\phi\right)-\delta_{ij}\left[\frac{1}{4}
F_{kl}F^{kl}+\frac{1}{2}D_k\phi^* D^k\phi \right].
\end{equation}
We examine the conditions (\ref{manton}), for real ($\phi=\phi^*$) configurations. Monopoles satisfy the Bogomolny equation $F_{ij}=\epsilon_{ijk}D_k\phi$, which implies
\begin{equation}
T_{ij}=(\delta_{ij}\delta_{kl}-\delta_{il}\delta_{kj})D_k\phi
D_l\phi +D_i\phi D_j\phi -\delta_{ij}\left[\frac{1}{2}(D_k\phi)^2
+\frac{1}{2}(D_k\phi)^2 \right]=0.
\end{equation}
We see that Manton's conditions are satisfied trivially for monopole solutions, as we discussed above this is because they saturate a BPS condition.

\subsection{Manton's conditions and electrodynamics beyond Maxwell}
It is interesting to consider Maxwellian electrodynamics in four
flat dimensions, extended to include higher order terms. The DBI
action is one example of such an action, but generically one can
write some arbitrary Lagrangian  $\cL (X,Y)$ as a function of the
two Lorentz-invariant combinations $X\equiv E^2-B^2$ and $Y\equiv
{\vec E}\cdot {\vec B}$. Using $E_i=F_{0i}$ and
$B_i=\frac{1}{2}\epsilon_{ilm}F_{lm}$, we can see that the energy
density is given by
\begin{align}
\cE = \frac{\delta\cL}{\delta F_{0i}}F_{0i}-\cL=2E^2\d_X\cL+Y\d_Y\cL-\cL~.
\end{align}
The stress tensor then becomes
\begin{align}
\Pi^i_{\phantom{i}j}&=\frac{\delta\cE}{\delta F_{0i}}F_{0j}+2\frac{\delta\cE}{\delta F_{ik}}F_{jk}-\cE\delta^i_j\nonumber\\
&=2\left(2E^2\d_X^2\cL+Y\d_X\d_Y\cL+\d_X\cL\right)E_iE_j+2\left(2E^2\d_X^2\cL+Y\d_X\d_Y\cL-\d_X\cL\right)B_iB_j\nonumber\\
&\quad+ \left( \cL-2X\d_X\cL-Y\d_Y\cL+2XY\d_X\d_Y\cL+Y^2\d_Y^2\cL-4E^2B^2\d_X^2\cL\right)\delta_{ij}~.
\end{align}
For Maxwell theory we have $\cL=\frac{1}{2}X$, which gives
\begin{align}
\Pi^i_{\phantom{i}j}=E_iE_j-B_iB_j-\frac{1}{2}\left( E^2-B^2\right)\delta_{ij}~.
\end{align}
$\Pi^i_{\ j}$ only vanishes for self-dual configurations
$E_i=\pm B_i$. Together with the equations of motion,
$\nabla\cdot\vec E=0$ and $\nabla\times \vec B=0$, one can see that
there are indeed no regular self-dual finite-energy static solutions.

These results may be useful for constructing valid Lagrangian with soliton solutions. For instance, if we have a Lagrangian of the form $\cL=Yf(X)$, the stress energy tensor becomes
\begin{align}
\Pi^i_{\ j} = 2\left( 2E^2Yf''+2Yf'\right)E_iE_j+4E^2Yf''B_iB_j-4E^2B^2Yf''\delta_{ij}~.
\end{align}
If one sets $E_i=0$, Manton's conditions are satisfied for any $B_i$, so in principle one could have solitons constructed entirely from magnetic fields.

\subsection{Restrictions on currents}\label{app:global}

Lorentz invariant theories that admit solitons have been thoroughly
investigated. In this subsection we survey these theories, focusing
on the space integrals of their conserved currents. We will check
whether these integrals indeed vanish, or equal non-vanishing
surface terms given by (\ref{intsurface}). We can classify the
currents as (i) topological currents (ii) currents associated with
global symmetries (iii) currents associated with local symmetries.

\paragraph{ Topological currents}

Topological currents are conserved without the use of equations of
motion. Their general structure in $d$ space dimensions  is of the
form \be J_\mu =
\epsilon_{\mu\nu_1...\nu_{d}}\tilde{J}^{\nu_1...\nu_{d}} 
\ee where  $\tilde{J}^{\nu_1...\nu_{d}}$ is a tensor of
degree $d$  composed of the underlying fields and their derivatives.

Consider now the spatial components of the current. If
$\tilde{J}^{\nu_1...\nu_{d}}$  is composed of scalar fields alone
(be they Abelian or non-Abelian),   it must include a time
derivative. For static configurations  the spatial components of the
current vanish automatically. When the theory also includes gauge
fields, for gauge-invariant topological currents  we can choose the
gauge $A_0=0$, and see that the spatial components of the current
again vanish trivially. For completeness, we list  here the
topological current of various models.
\begin{itemize}
\item For the Sine-Gordon model \cite{Coleman}, the current is $J_\mu= \epsilon_{\mu\nu}\pa^\nu\phi$ where $\phi$ is a scalar field.
\item The baryon number in bosonized 1+1 dimensional QCD \cite{Date:1986xe}
is given by $J_\mu = \epsilon_{\mu\nu}Tr[ g^{-1}\pa^\nu g]$ where
$g\in U(N_f)$ see the next subsection.
\item In 3+1 dimensions the current
of the non-Abelian $SU(2)$ magnetic monopole \cite{'tHooft:1974qc,Polyakov:1974ek} is $J_\mu=
\epsilon_{\mu\nu\rho\sigma}\epsilon_{abc}\pa^\nu\phi^a\pa^\rho\phi^b\pa^\sigma\phi^c$
where $\phi^a$ is a scalar field in the adjoint representation  of
the group.
\item For the Skyrme model \cite{Skyrme:1962vh}, the current is $J_\mu=
\frac{\epsilon_{\mu\nu\rho\sigma}}{24\pi^2}Tr[L^\nu L^\rho
L^\sigma]$, where $L_\mu= u^{-1}\pa_\mu u$ and where $u$ is the
Skyrme field.
\item The topological current for the four-dimensional gauge
instanton is $J_\mu=
\frac{\epsilon_{\mu\nu\rho\sigma}}{16\pi^2}Tr[A^\nu\pa^\rho
A^\sigma+ \frac{2}{3} A_\nu A_\rho A_\sigma]$ where $A_\mu$ is the
corresponding non-Abelian gauge field.
\item In five-dimensional gauge
theories, the topological current is given by
$J= *F\wedge F$.
\end{itemize}

\paragraph{ Global currents}

Next we examine  currents associated with  global continuous
symmetries in two cases where the key player is a scalar field,
which is a group element (1)  of the 1+1 dimensional bosonized
multi-flavor QCD\cite{Date:1986xe}, and (2) of the  Skyrme model in
3+1 dimensions \cite{Skyrme:1962vh}.
\begin{enumerate}
\item
Though the sine-Gordon model in 1+1 dimensions has no global
conserved currents, its non-Abelian generalization, 1+1 dimensional
bosonized multi-flavor QCD\cite{Date:1986xe}, does. In the strong
coupling limit one can integrate over the color degrees of freedom,
leaving a theory invariant under a $U(N_f)$ global symmetry of the
form \be\label{mfqcdaction} S= N_c S_{WZW}(g) + m^2 \int d^2 x \Tr
[g+ g^\dagger] \ee where $N_c$ is the number of colors, $S_{WZW}$ is
the WZW action, $g$ is a group element  $g\in U(N_f)$. $m$ is a mass
parameter which depends on the coupling and the quark mass, as given
in \cite{Date:1986xe}. This ``mass term'' is essential for the
existence of stable soliton solutions, which describe the baryons of
the theory. This is thus a 1+1 dimensional analog of the Skyrme
model. The vector and  axial flavor currents  associated with the
WZW action take the form \be\label{mfQCDcurrents}
J_\mu^{\frac{V}{A}}= \frac{iN_c}{8\pi}[(g^{-1}\pa_\mu g\pm g\pa_\mu
g^{-1})+\epsilon_{\mu\nu}(g^{-1}\pa^\nu g\mp g\pa^\nu g^{-1})]~. \ee
For a static  configuration the space component of the currents gets
a contribution only from the  first term. The classical soliton
solution has the structure \be g_0(x)=\, {\rm diag}\,\left (
1,...1,e^{-i\sqrt{\frac{4\pi}{N_c}}\varphi(x)}\right) \ee
 where $\varphi(x)$ is a solution of the sine-Gordon equation $\pa_x^2\varphi-\mu^2\sin(\beta\varphi)=0$ with $\beta= \sqrt{\frac{4\pi}{N_c}}$ and $\mu^2= 2\sqrt{\frac{4\pi}{N_c}} m^2$.
 It is straightforward to see that for this configuration the space component of the vector current vanishes.
 The spatial part of the axial current does not vanish: indeed, the second term  in the action \eqref{mfqcdaction}
   is not invariant  axial transformations.

\item
In the four dimensional theory of ``bosonized" baryons in the Skyrme
model, integrals over the flavor currents are more interesting. The
model of two flavors, without a mass term,  is invariant under both
the $SU(2)$ vector and axial flavor global symmetries. No mass term
is needed to have soliton solutions, since the Skyrme term plays the
role of stabilizing the solitons. The axial current is proportional
to the first term of (\ref{mfQCDcurrents}) plus higher derivative
corrections from the Skyrme term. Recall that the  two-flavor Skyrme
model does not include a  WZ term. Unlike the 1+1-d case, here the
space components of the axial current do {\it not} vanish. The non-Abelian (axial) current was found in
\cite{Adkins:1983ya} to be
 \be
 {J_i^a}=\frac14F_\pi\frac{B}{r^3}[(\tau_i-3\vec\tau\cdot\hat x \hat x_i )\tau^a] +...
 \ee
 so that
 \be
 \int d^3x {J_i^a}= -\frac23 F_\pi^3 B \pi \delta_i^a
 \ee
 where $B= \frac{8.6}{e^2F_\pi^2}$ ($\frac{1}{32 e^2}$  is the coefficient of the Skyrme term).
 Hence the volume integral of the space component of the axial current
 takes the form of (\ref{intsurface}), yielding the Skyrme model prediction for the axial coupling.
\end{enumerate}

\paragraph{Currents associated with local symmetries}

Let us now turn to gauge theories, which have conserved currents
that follow from local symmetries. Gauge transformations leave the
action invariant, so we perform transformations keeping the gauge
fields fixed. For non-Abelian symmetries the current is not just
conserved, but covariantly conserved. The variation of the energy is
still a total derivative, however:
\begin{equation}
\delta_\theta E=\int d^d x\, D_i \theta^A J_A^i=\int d^d x\,\left[\partial_i \left(\theta^A J_A^i\right)-\theta^A D_i J_A^i\right]=\oint d^{d-1} \sigma\, n_i\theta^A J_A^i.
\end{equation}
We discuss first the Abelian Higgs model \cite{Nielsen:1973cs,de Vega:1976mi} and then the model of the
$SU(2)$ non-Abelian magnetic monopole  \cite{'tHooft:1974qc,Polyakov:1974ek}. In both two cases there are scalar
fields that do not vanish at infinity, so the transformation we
perform would in principle affect to the boundary conditions on the
fields. However, it turns out that in these cases the constraints are
also satisfied.
\begin{enumerate}
\item
The Abelian Higgs model is defined by the following Lagrangian density
\be
{\cal L}= -\frac{1}{4} F^{\mu\nu}
F_{\mu\nu} + \frac12 |(\pa_\mu +ie A_\mu)\phi|^2   - c_2|\phi|^2 - c_4(|\phi|^2)^2
\ee
where $\phi$ is a complex scalar field and $A_\mu$ is an Abelian gauge field.
 The corresponding equations of motion admit soliton solutions in the form of
vortices \cite{Nielsen:1973cs}. The vortex has formally infinite
energy because it is extended in the $z$ direction, but at each
fixed value of $z$ the energy density per unit length is finite.  We
parameterize the complex scalar field and the gauge field in polar
coordinates $\rho$ and $\varphi$ (there is no dependence on $z$) as
\be\label{abhigsol} \phi(\vec x)= e^{-in\varphi}f(\rho) \qquad \vec
A(\vec x) =-\frac{A(\rho)}{\rho} \hat \varphi, \ee where $n$ is an
integer. It was shown in \cite{deVega:1976mi} that for $c_4=e^2/8$
the soliton is BPS,
 hence the components of the stress  tensor vanish $T_{\rho\rho}=T_{\varphi\varphi}=0$  and (\ref{manton}) is automatically satisfied.
 The model admits also an Abelian conserved
current \be \vec{J} = i[\phi^*\vec{D}\phi- (\vec{D}\phi^*)\phi]=
2\left(n+A(\rho)\right) \frac{(f(\rho))^2}{\rho}\hat \varphi~. \ee
At large values of the radial coordinate  $f$ and $A$ go to
constants, $ \lim_{\rho\rightarrow \infty}f(\rho)=
\sqrt{\frac{c_2}{2 c_4}}$ and $\lim_{\rho\rightarrow \infty}A(\rho)=
-n$. The current indeed vanishes at infinity.

We now examine the conditions \eqref{currentcond} for a volume whose
boundary is a cylinder of radius $\rho=R$ and with caps at $z=\pm
L$.\footnote{We can repeat the same analysis in 2+1 dimensions, in
which case the boundary is just a circle in the plane.} They take
the  form
 \be
 \int d^3 x \pa_i( J^i \theta) = \lim_{R,L\to \infty}\oint
  d\varphi\left[ \int_{-L}^L dz  \left.\theta J_\rho\right|_{\rho=R}+ \int_0^R \rho d\rho \left(\left.\theta J_z\right|_{z=L}-\left.
  \theta J_z\right|_{z=-L}\right)\right]=0~.
\ee These conditions are automatically satisfied as $J_\rho=J_z=0$.

\item
We can also check the conditions \eqref{currentcond} for the 't
Hooft-Polyakov\cite{Polyakov:1974ek,'tHooft:1974qc} monopole. The
components of the stress tensor vanish because the BPS condition is
saturated.

 The system includes $SO(3)$ gauge fields
interacting with an iso-vector scalar field, with Lagrangian density
\be {\cal L}= -\frac{1}{4} F^{\mu\nu}_a F^a_{\mu\nu} + \frac12 D_\mu
\vec\phi\cdot D^\mu\vec\phi  - \frac14 \lambda( |\phi|^2 - v^2 )^2
\ee where $a=1,2,3$. The current associated with the $SO(3)$ global
symmetry is given by \be J^a_\mu = \epsilon^a_{bc} (D_\mu\phi)^b
\phi^c \qquad  (D_\mu\phi)^a = \pa_\mu\phi^a -e \epsilon^a_{bc}
A_\mu^b\phi^c \ee The corresponding equations of motion read \be
(D_\nu F^{\nu\mu})^a = J_\mu^a \qquad (D^\mu D_\mu\phi)_a =
-\lambda\phi^a( |\phi^2|-v^2)^2 \ee The relevant ansatz for
solutions of the scalar and gauge fields reads \be \phi^a(\vec r) =
H( evr)\frac{x^a}{er^2} \qquad A_i^{a}(\vec x) = -\epsilon^{a}_{\
ij} \frac{x^j}{er^2}[1-K(evr)] \ee where the functions $H$ and $K$,
which are determined by solving the equations of motion, behave
asymptotical as $H\sim evr$ and  $K\sim evre^{-evr}$. Substituting
the ansatz into the current we find \be J^a_i = \left(\frac{H(
evr)}{er^2}\right)^2 K( evr) \epsilon^a_{\ ij}x^j~. \ee It is thus
obvious that \be \int d^3 x J^a_i = \lim_{r\to \infty} \oint
d\Omega_2\, x^i \hat{x}^k J^a_k=0 \ee in accordance with
(\ref{currentcond}), as the projection of the current on the radial
direction vanishes.

\end{enumerate}


\section{Conditions for brane and string actions}

In the previous section, we illustrated the use of DC in known
examples and some simple extensions. Even for these relatively
simple models, finding solitonic solutions can be highly
non-trivial, and is usually the result of a clever ansatz. For
D-brane systems, in which the DBI action generates an infinite
number of interaction terms, the task can be that much more
difficult. The DC represent a relatively simple test which may
exclude the existence of solitons in large classes of systems. While
satisfying the DC is necessary but not sufficient for soliton
solutions, they may also help to construct ans\"{a}tze for new
static solutions.

We note again that one should be careful when dealing with finite
volume configurations, as the conditions we impose are based on
transformations that generically do not leave boundary conditions
invariant. As mentioned above, one can only compare the energies of
configurations satisfying the same boundary conditions.

The low-energy dynamics of D$p$-branes in a weakly curved gravitational background are given by the Dirac-Born-Infeld (DBI) action,
\begin{equation}
S_{DBI}=-T_p\int d^{p+1}x e^{-\phi}\sqrt{-\det\left(g_{\mu\nu}+2\pi\alpha' F_{\mu\nu} +B_{\mu\nu}\right)}.
\end{equation}
$T_p$ is the $Dp$ brane tension, $\phi$ is the background dilaton
field, $B_{\mu\nu}$ is the pullback of the bulk Neveu-Schwarz (NS)
two-form,  $F_{\mu\nu}$ is the field strength of the Abelian gauge
field on the brane, and $g_{\mu\nu}$ is the pullback of the
background metric to the brane worldvolume. In the following we will
use Greek letters ($\mu,\nu,\dots$) to denote worldvolume
coordinates and capital Latin letters ($M,N,\dots$) to denote bulk
coordinates and indices. Later we will also use lower case Latin
indices ($i,j,\dots)$ for the spatial coordinates on the brane
worldvolume. The metric has Lorentzian signature and we continue
to use a mostly plus convention. This action is valid only to
leading
 order
 in an expansion involving gradients of the gauge field and higher
 curvature corrections to the background, with $\alpha'$ setting the scale of the expansion.

D-branes also carry charges that couple to Ramond-Ramond (RR) forms,
as described by the Wess-Zumino (WZ) action
\begin{equation}
S_{WZ}=T_p\int d^{p+1}x \sum_k C_k \wedge e^{2\pi\alpha' F+B},
\end{equation}
where $C_k$ is the pullback of the $k$-form RR potentials.

A very similar semiclassical description exists for fundamental
strings, via the Nambu-Goto (NG) action:
\begin{equation}
S_{NG}=-\frac{1}{2\pi\alpha'}\int d^2 x \sqrt{-\det\left(g_{\mu\nu}\right)}.
\end{equation}
The string is charged under the NS two-form, with a WZ action
\begin{equation}
S_{B}=\frac{1}{2\pi\alpha'}\int d^2 x B.
\end{equation}
Given the similarity between D-brane effective actions and string
actions, we can consider them within the same framework.

The branes' embedding in a $D$-dimensional background is described
by a set of scalars living on its worldvolume $X^M(x)$,
$M=1,\cdots,D$. The scalars appear in the pullback of the metric,
\begin{equation}
g_{\mu\nu}=G_{MN}\partial_\mu X^M\partial_\nu X^N,
\end{equation}
where $G_{MN}$ is the metric of the background and can be a
functional of the scalar fields $X^M$ when evaluated on the brane
worldvolume. The DBI+WZ actions together constitute the action of a
$p+1$-dimensional theory including the scalar fields $X^M$ and the
gauge fields $F_{\mu\nu}$.\footnote{Of course, there will be no
internal gauge fields when we restrict to the case of strings
described by the NG action.} For infinitely extended configurations,
there should be no finite energy static configurations if there are
no background forms, since nontrivial excitations of the worldvolume
scalars increase the energy. Background forms, however, can  act as
chemical potentials that decrease the energy density of the branes.
We can tune the potentials so that trivial configurations have zero
energy, which for the simplest cases  amounts to subtracting off the
ground state energy, as one  expects that the restriction on the
existence of static soliton solutions still applies when the forms
are turned off.

Before deriving Manton's conditions in detail, we address
a subtlety of these types of actions.
The brane action is invariant under worldvolume diffeomorphisms, which na{\"\i}vely would
lead to empty conditions when one applies Derrick's scaling of the coordinates. Let us illustrate this for the simplest case of the NG action.
We assume that there is a static solution to the equations of motion $X^M(x)$, $\partial_t X^M=0$ and evaluate the energy for $X^M(\lambda x)$:
\begin{align}
\notag E(\lambda) &=-S_{NG}=\frac{1}{2\pi\alpha'}\int dt dx \sqrt{-G_{MN}\partial_x X^M(\lambda x) \partial_x X^N(\lambda x)} \\
&=\frac{1}{2\pi\alpha'}\int dt dx' \lambda^{-1} \sqrt{- \lambda^2 G_{MN}\partial_{x'} X^M(x') \partial_{x'} X^N(x')}=E(1).
\end{align}
In other words, the scaling is eliminated by the diffeomorphism invariance of the action. In order to find non-trivial conditions, we must gauge-fix the diffeomorphism invariance first.  We will choose the standard static gauge everywhere, with the scalar fields $X^M$ split into those pointing along the worldvolume directions $X^\mu$ ($\mu=0,1,\dots,p$), and those pointing along the transverse directions $Y^a$ ($a=p+1,\dots,D-1$):
\begin{align}
\notag X^\mu(x)=& \ x^\mu\\
X^a(x)=& \ Y^a(x)
\end{align}
This choice may not be suitable for all possible situations, for
instance if there are solutions that join two brane sheets by a
throat \cite{Gibbons:1997xz,Callan:1997kz}. These will be
multivalued in the static gauge and can evade the na\"{i}ve
constraints. We will ignore this issue in what follows.

We impose some other simplifying assumptions as well, which
nevertheless include many physically interesting cases. We impose
\begin{itemize}
\item time and space translation invariance of the worldvolume action. This implies that the background metric, dilaton and RR forms will be independent of the directions along the brane.
\item that the solutions be truly static, meaning that $\partial_t Y^a =0$. One could relax this condition keeping invariance under time translations of the action if $\partial_t Y^a$ is constant and the background metric is independent of $Y^a$.
\end{itemize}
Note that in order to apply the DC we need the energy to be a
conserved quantity: time translation invariance is thus a necessary
requirement. When we come to nontrivial gauge field configurations,
we will include static electric fields, however. These are allowed
because the actions we study are gauge invariant.

The outline of the rest of this section is as follows: first we
study Derrick-type DC conditions on brane actions without gauge
fields and WZ terms,\footnote{We will always, however, include a
term which sets the ground state energy of the system to zero.} then
we derive the general form of the DC, including gauge fields;
finally we apply the conditions to the lower dimensional cases (D1,
D2 and D3) including also WZ terms, and consider some interesting
examples.

\subsection{Scaling constraints in brane actions with scalars}

We start by assuming that there is a static solution to the
equations of motion $Y^a(\{x^i\})$ depending only on the spatial
coordinates, with no gauge fields turned on. The pullback of the
metric in the static gauge is
\begin{align}
g_{00}=G_{00}, \ \ g_{0i}=G_{0i}, \ \ g_{ij}=G_{ij}+G_{ab}\partial_i Y^a\partial_j Y^b.
\end{align}
Let us now evaluate the energy on some field configuration $Y^a(\{
\lambda_i x^i\})$ (no summation). After re-scaling the spatial
coordinates as $x^i\rightarrow \lambda^ix^{i}$ (no sum) we will have
\begin{equation}\label{eq:EDBIscalars}
E(\{\lambda_i\})=T_p\int d^p x\,e^{-\epsilon\phi}\left(\prod_{i=1}^p \lambda_i^{-1}\right)\left[ \sqrt{-\det(g[\{\lambda_i\}]})- \sqrt{-\det G}\right].
\end{equation}
For a D-brane $\epsilon=1$ and $T_p$ is the D-brane tension. For a
string $\epsilon=0$, $p=1$ and $T_p$ is the string tension. The last
term is due to a background form and is chosen such that for
$\partial_i Y^a=0$ the energy is zero.

Derrick's theorem does \textit{not} explicitly
forbid the existence of scalar-field solitonic solutions in the DBI
action.
If a static solution indeed exists and is stable, the energy should be stationary as a function of the $\lambda_i$:
\begin{equation}\label{derrickbrane}
0=\frac{dE}{d\lambda_i}=-T_p\int d^p x\,e^{-\epsilon\phi}\left(\sqrt{-\det g}-\sqrt{-\det G} -\sqrt{-\det g}\sum_l g^{il} (g_{il}-G_{il})\right)~,\end{equation}
where we have used
\begin{equation}
\frac{d g_{kl}}{d\lambda_i}=2\delta_k^i\lambda_l G_{ab}\partial_i Y^a \partial_l Y^b=2\delta_k^i\lambda_l(g_{il}-G_{il}).
\end{equation}
For the trivial configuration $g_{ij}=G_{ij}$, and Derrick's conditions are clearly satisfied, but, as discussed in Appendix \ref{Dbranesol}, solitons cannot be excluded on general grounds. However, using conditions like \eqref{manton} we can make concrete statements about some special cases.

Consider Dp-branes with arbitrary $p$, such that the pullback fields
$Y^a$ depend only on a single coordinate, which we call $x$.
Assuming $G_{0i}=0$, we can show that solitons of this type are not
allowed from the DBI action alone. The regularized energy is given
by
\begin{align}
E=T_p\int d^px e^{\epsilon\phi}\left[\sqrt{-\det g}-\sqrt{-\det
G}\right]~,
\end{align}
where for this simple case
\begin{align}
\det g =  \left( 1+{G}^{xx}G_{ab}\d_x Y^a\d_x Y^b\right)\det G
\end{align}
so
\begin{align}
E=T_p\int d^px e^{\epsilon\phi}\left\{ \sqrt{(-\det
G)\left[1+{G}^{xx}G_{ab}\d_x Y^a\d_x Y^b\right]}-\sqrt{-\det
G}\right\}~.
\end{align}
Taking $x\rightarrow\lambda x$ in the solution (and leaving the
other coordinates fixed) we find
\begin{align}
\left.\frac{dE}{d\lambda}\right|_{\lambda=1}=0=T_p\int d^px \
\sqrt{\frac{-\det G}{1+{G}^{xx}G_{ab}\d_x Y^a\d_x Y^b}}\left\{
\sqrt{1+{G}^{xx}G_{ab}\d_x Y^a\d_x Y^b}-1 \right\}~.
\end{align}
For nontrivial $\d_xY^a,$ the integrand must be greater than zero,
and the condition \eqref{manton} cannot be satisfied! Note that this
is the case for any brane dimension, even $p=1$. We provide more
detail for a particular example in \S \ref{sec:probebranes}.

\subsection{DC in brane actions with scalars and gauge fields}

We now consider the more complicated case of D-branes with gauge
fields and scalars, deriving general conditions based on both
scaling and shear deformations.

When electric fields are turned on, the energy is the Legendre
transform of the Lagrangian,
\begin{equation}
E_{\rm can}=\int d^{p}x\,\frac{\delta S_{DBI}}{\delta \partial_0 A_i}\partial_0 A_i- S_{DBI}.
\end{equation}
Since the DBI action is gauge invariant one can also write the energy as
\begin{align}
\notag E_{\rm can} &=\int d^{p}x\,\frac{\delta S_{DBI}}{\delta F_{0i}}(F_{0i}+\partial_i A_0)- S_{DBI}=\int d^{p+1}x\,\frac{\delta S_{DBI}}{\delta F_{0i}}F_{0i}- S_{DBI}\\
&+\int d^{p+1} x \partial_i\left(A_0 \frac{\delta S_{DBI}}{\delta F_{0i}}\right).
\end{align}
In the last term we have used Gauss's law
\begin{equation}
\partial_i\left(\frac{\delta S_{DBI}}{\delta F_{0i}}\right)=0.
\end{equation}
We can eliminate it by adding the improvement term
\begin{equation}
\Psi^{i0}_{\ \ 0}=-A_0\frac{\delta S_{DBI}}{\delta F_{0i}}.
\end{equation}
Note that for static solutions $\partial_0 \Psi^{0i}_{\ \
0}=-\partial_0 \Psi^{i0}_{\ \ 0}=0$, so the improvement term does
not affect the $T^i_{\ 0}$ components of the energy-momentum tensor.
The energy functional takes the expected gauge-invariant form
\begin{equation}
E_{DBI}=\int d^{p+1}x\,\frac{\delta S_{DBI}}{\delta F_{0i}}F_{0i}- S_{DBI}.
\end{equation}
It will be convenient to write the argument of the square root factorizing the pullback of the metric from the gauge fields
\begin{equation}
g_{\mu\nu}+2\pi\alpha' F_{\mu\nu}=g_{\alpha\nu} M_{\mu}^{\ \alpha},
\end{equation}
where
\begin{align}
M_\mu^{\phantom{\mu}\nu}=\delta_\mu^{\ \nu}+2\pi\alpha'F_\mu^{\phantom{\mu}\nu}~.
\end{align}
The determinant factorizes as
\begin{equation}
\det\left(g_{\mu\nu}+2\pi\alpha' F_{\mu\nu} \right)=\det g \,\det M.
\end{equation}
To find the variation of $\det M$ with respect to the gauge fields we use the formula
\begin{align}
\det M=\frac{1}{(p+1)!}\epsilon^{\mu_1\dots
\mu_{p+1}}\epsilon_{\nu_1\dots
\nu_{p+1}}M_{\mu_1}^{\phantom{\mu_1}\nu_1}M_{\mu_2}^{\phantom{\mu_2}\nu_2}\dots
M_{\mu_{p+1}}^{\phantom{\mu_{p+1}}\nu_{p+1}}~,
\end{align}
where the $\epsilon^{\mu_1\cdots \mu_{p+1}}=\epsilon_{\mu_1\cdots
\mu_{p+1}}$ are Levi-Civita symbols. Then,
\begin{align}\label{variationF}
\frac{\delta S_{DBI}}{\delta F_{\mu\nu}}=-\frac{T_p}{2}
e^{-\phi}\sqrt{\frac{-\det g}{\det M}}\frac{1}{p!}\epsilon^{\mu\mu_2\dots
\mu_{p+1}}\epsilon_{\nu_1\dots
\nu_{p+1}}(2\pi\alpha')g^{\nu\nu_1}M_{\mu_2}^{\phantom{\mu_2}\nu_2}\dots
M_{\mu_{p+1}}^{\phantom{\mu_{p+1}}\nu_{p+1}}~.
\end{align}
The energy becomes
\begin{equation}
E_{DBI}=T_p \int d^{p}x e^{-\phi}\sqrt{\frac{-\det g}{\det M}}\left[\det M-\frac{1}{2p!} \epsilon^{0\mu_2\dots
\mu_{p+1}}\epsilon_{\nu_1\dots
\nu_{p+1}}(M_0^{\ \nu_1}-\delta_0^{\ \nu_1})M_{\mu_2}^{\phantom{\mu_2}\nu_2}\dots
M_{\mu_{p+1}}^{\phantom{\mu_{p+1}}\nu_{p+1}} \right]
\end{equation}
Defining $\epsilon^{i_1\cdots i_{p}}=\epsilon^{0 i_1\cdots i_{p}}$, we can further simplify this expression to
\begin{equation}
E_{DBI}=T_p \int d^{p}x e^{-\phi}\sqrt{\frac{-\det g}{\det M}}\left[\det M-\frac{1}{2} \left(\det M-\det\hat{M}\right)\right] .
\end{equation}
Following the notation of Appendix \ref{Dbranesol}, a hatted matrix
only has indices in the spatial directions, so
$\det\hat{M}=\det(M_i^{\ j})$. Finally, we can write the energy as
\begin{equation}
E_{DBI}=\frac{T_p}{2} \int d^{p}x e^{-\phi}\sqrt{-\det g\det M}\left[1+\frac{\det\hat{M}}{\det M}\right] .
\end{equation}
If the electric field is zero $\det M=\det\hat{M}$ and one recovers the result that the energy is minus the DBI action.

Assuming there exists some finite energy solution to the equations
of motion, $(Y^a(\{x^i\}), F_{\mu\nu}(\{x^i\}))$, the energy should
be stationary under deformations of the solution. A general
infinitesimal deformation is characterized by a vector $\xi^i(x)$
linear in the coordinates. The deformed configurations are
\begin{equation}
Y^a(\{x^i+\xi^i\}), \ \ F^\xi_{\mu\nu}=F_{\mu\nu}(\{x^i+\xi^i\})+\partial_\mu \xi^\alpha F_{\alpha\nu}(\{x^i+\xi^i\})+\partial_\nu \xi^\alpha F_{\mu \alpha}(\{x^i+\xi^i\}).
\end{equation}
In terms of electric and magnetic fields
\begin{align}
F^\xi_{0i} &=F_{0i}(\{x^i+\xi^i\})+\partial_i \xi^j F_{0j}(\{x^i+\xi^i\}), \\  F^\xi_{ij} &=F_{ij}(\{x^i+\xi^i\})+\partial_i \xi^k F_{kj}(\{x^i+\xi^i\})+\partial_j \xi^k F_{i k}(\{x^i+\xi^i \}).
\end{align}
We introduce these expressions in the energy and perform a change of
coordinates $x^i\to x^i-\xi^i(x)$. The change in the energy is
\begin{equation}
\delta_\xi E=\int d^p x\,\delta_\xi\cE=-\int d^p x\, \Pi^i_{\ j}\partial_i \xi^j,
\end{equation}
where we have defined the stress tensor as
\begin{equation}
\Pi^i_{\ j}=\frac{\delta E_{DBI}}{\delta (\partial_i Y^a)}\partial_j Y^a+\frac{\delta E_{DBI}}{\delta F_{0i}}F_{0j}+2\frac{\delta E_{DBI}}{\delta F_{ik}}F_{jk}-\delta^i_{\ j}\cE.
\end{equation}
We will now compute each term in this expression. We will use the
fact that
\begin{equation}
\frac{\delta E_{DBI}}{\delta (\partial_i Y^a)}=-\frac{\delta E_{DBI}}{\delta g^{\mu\nu}}g^{\mu\alpha} g^{\nu\beta}\frac{\delta g_{\alpha\beta}}{\delta (\partial_i Y^a)}=-\frac{\delta E_{DBI}}{\delta g^{\mu\nu}}(g^{\mu k} g^{\nu i}+g^{\mu i} g^{\nu k})G_{ab}\partial_k Y^b.
\end{equation}
We will compute the variation of $E_{DBI}$ with respect to the
metric further down, but first consider the terms involving the
gauge fields. In order to vary with respect to the electric field,
note that $\det g$ is independent of $F_{0i}$, so the only
contributions come from the variation of $\det M$ and $\det\hat{M}$.
We can simplify the result by using
\begin{equation}
\frac{\delta }{\delta F_{0i}}\left( \frac{1}{\sqrt{\det M}}\right)=-\frac{1}{\det M}\frac{\delta \sqrt{\det M}}{\delta F_{0i}}.
\end{equation}
Then, the variation of the energy is
\begin{equation}
\frac{\delta E_{DBI}}{\delta F_{0i}}=-\frac{1}{2}\frac{\delta S_{DBI}}{\delta F_{0i}}\left[ 1-\frac{\det\hat{M}}{\det M}\right]+\frac{T_p}{2}e^{-\phi}\sqrt{\frac{-\det g}{\det M}}\frac{\delta \det\hat{M}}{\delta F_{0i}}.
\end{equation}
where the variation of the DBI action was given in formula
\eqref{variationF}. For the magnetic components the situation is
similar,
\begin{equation}
\frac{\delta E_{DBI}}{\delta F_{ik}}=-\frac{1}{2}\frac{\delta S_{DBI}}{\delta F_{ik}}\left[ 1-\frac{\det\hat{M}}{\det M}\right]+\frac{T_p}{2}e^{-\phi}\sqrt{\frac{-\det g}{\det M}}\frac{\delta \det\hat{M}}{\delta F_{ik}}.
\end{equation}

We will now compute the missing pieces, starting with the variation of the energy with respect to the metric
\begin{equation}
\frac{\delta E_{DBI}}{\delta g^{\mu\nu}}=-\frac{1}{2}\cE g_{\mu\nu}+\frac{T_p}{4}e^{-\phi}\sqrt{\frac{-\det g}{\det M}}\left[1-\frac{\det\hat{M}}{\det M}\right]\frac{\delta \det M}{\delta g^{\mu\nu}}+\frac{T_p}{2}e^{-\phi}\sqrt{\frac{-\det g}{\det M}}\frac{\delta \det \hat{M}}{\delta g^{\mu\nu}}.
\end{equation}
Note that $M$ depends on $F_{\mu}^{\ \nu}=g^{\nu\alpha}F_{\mu\alpha}$, then
\begin{equation}
\frac{\delta \det M}{\delta g^{\mu\nu}}=g_{\mu\beta} F_{\alpha\nu}\frac{\delta \det M}{\delta F_{\alpha\beta}}.
\end{equation}
The same formulas can be used substituting $\det M$ by $\det\hat{M}$, since $\hat{M}$ depends on $F_i^{\ j}=g^{j\alpha} F_{i\alpha}$:
\begin{equation}
\frac{\delta \det\hat{M}}{\delta g^{\mu\nu}}=\frac{\delta \det\hat{M}}{\delta F_i^{\  j}} F_{i\nu} \delta^{\ j}_{\mu}, \ \ \frac{\delta \det\hat{M}}{\delta F_{\alpha\beta}}=\frac{\delta \det\hat{M}}{\delta F_i^{\ j}}\delta^{\ \alpha}_i g^{\beta j}.
\end{equation}
In order to simplify the expressions we will use the relation
\begin{equation}
\frac{T_p}{2}e^{-\phi}\sqrt{\frac{-\det g}{\det M}}\frac{\delta \det M}{\delta F_{\alpha\beta}}=-\frac{\delta S_{DBI}}{\delta F_{\alpha\beta}}.
\end{equation}
Finally, adding all the terms we have the stress tensor
\begin{align}\label{stressDBI}
\Pi^i_{\ j} &=\cE_{DBI}\left(g^{ik}G_{ab}\partial_k Y^a\partial_j Y^b-\delta^i_{\ j} \right)+\delta^i_j\cE_{DBI}^0\\
\notag &+\frac{1}{2}\left[\frac{\delta S_{DBI}}{\delta F_{0l}}\left[1-\frac{\det\hat{M}}{\det M}\right]+\frac{\cL_{DBI}}{\det M}\frac{\delta \det\hat{M}}{\delta F_{0l}}\right]\left((F_0^{\ i}\delta_l^k+F_0^{\ k}\delta_l^i)G_{ab}\partial_k Y^a\partial_j Y^b-\delta^i_l F_{0j} \right)\\
\notag &+\frac{1}{2}\left[\frac{\delta S_{DBI}}{\delta
F_{nl}}\left[1-\frac{\det\hat{M}}{\det
M}\right]+\frac{\cL_{DBI}}{\det M}\frac{\delta \det\hat{M}}{\delta
F_{nl}}\right]\left((F_n^{\ i}\delta_l^k+F_n^{\
k}\delta_l^i)G_{ab}\partial_k Y^a\partial_j Y^b-2\delta^i_n F_{jl}
\right)
\end{align}
where the index on $F_i^l$ in the above are raised using the
pullback metric, $g$. In the first line of this expression we have
included the term $\cE_{DBI}^0\equiv \cE_{DBI}(A_\mu=0,Y^a=0)$, the
energy of the trivial configuration. This guarantees that the net
energy density, $\cE_{DBI}-\cE_{DBI}^0$, is finite.  When
Ramond-Ramond fluxes are included in Dp-brane action, the integral
of the $p$-form flux associated with the brane assumes on this role,
so one must be careful not to double-count its contribution (for
instance by always normalizing the trivial configuration to have
zero energy).

\paragraph{Stress tensor for BPS configurations}\label{bpsbrane}

As we discussed previously, solitons that saturate a BPS bound
should have vanishing (spatial) stress tensor components. We check
this now for a few examples.

Let us first examine the case where the gauge fields are turned off and $G_{0i}=0$. We find
\begin{equation}
\Pi^i_{\  j}=\cE_{DBI}(g^{ik}(g_{kj}-G_{kj})-\delta^i_{\  j})+\delta^i_j\cE_{DBI}^0=\delta^i_j\cE_{DBI}^0-\cE_{DBI}g^{ik}G_{kj}.
\end{equation}
It is easy to see that the components of the stress tensor vanish
for the Abelian vortices constructed in \cite{Gauntlett:1997ss}. The
action is that of the M2 brane in flat space, which is the same as
the $2+1$ dimensional DBI without the dilaton factor. We pick two
directions along the brane, $x^i$ with $i=1,2$, and two directions
transverse to the brane  $Y^1=X$, $Y^2=Y$. We then impose the BPS
condition
\begin{equation}\label{vortexBPS}
\partial_i X=\pm \epsilon_{ij}\partial_j Y.
\end{equation}
In this simple case, if we factor out the tension of the M2 brane,
$\cE^0=1$, $\cE=(1+(\partial Y)^2)$ and $g_{ij}=(1+(\partial
Y)^2)\delta_{ij}$. Then,
\begin{equation}
\Pi^i_{\  j}=\delta^i_j-(1+(\partial Y)^2)\frac{\delta^{ik}}{(1+(\partial Y)^2)}\delta_{kj}=0.
\end{equation}

Another example of a BPS configuration is an instanton in the
D4-brane worldvolume. Here we allow magnetic but not electric fields
or scalars to be turned on. Using $\det\hat{M}=\det M$, the
components of the stress tensor  (in units of the tension) become
\begin{equation}
\Pi^i_{\ j} =\delta^i_j(1-\cE_{DBI})+\frac{\cE_{DBI}}{\det \hat{M}}\frac{\delta \det\hat{M}}{\delta F_{ik}}F_{jk}.
\end{equation}
When the (anti) self-duality condition $F_{ij}=\pm \frac{1}{2}\epsilon_{ijkl} F_{kl}$ is satisfied,
\begin{equation}
\frac{\delta \det\hat{M}}{\delta F_{ik}}= \sqrt{\det \hat{M}} F^{ik},
\end{equation}
and the energy density becomes $\cE_{DBI}=\sqrt{\det
\hat{M}}=1+\frac{1}{4}F_{ij}F^{ij}$. Then,
\begin{equation}
\Pi^i_{\ j} =-\frac{1}{4}\delta^i_j F_{kl}F^{kl}+F^{ik}F_{jk}.
\end{equation}
This is the stress tensor for Maxwell theory. We can also write it
as
\begin{equation}
\Pi^i_{\ j}=\frac{1}{2}\left(F^{ik}-\frac{1}{2}\epsilon^{ikln}F_{ln} \right)\left(F_{jk}+\frac{1}{2}\epsilon_{jkln}F_{ln} \right)-\frac{1}{4}\left(\epsilon_{jkln} F^{ik} F^{ln}-\epsilon^{ikln} F_{jk} F_{ln} \right).
\end{equation}
When the (anti) self-duality condition is satisfied it is easy to
check that this vanishes exactly.

\subsubsection{Adding Wess-Zumino terms}

In order to make the energy of the brane finite, we can add an
additional term from the WZ action which precisely cancels the
energy density of trivial configurations. This term comes from the
coupling of a Dp-brane to the RR $(p+1)$-form potential, and
contains a variety of terms involving gauge fields as well:
\begin{equation}
S_{WZ}=T_p \int_{Dp} C_{p+1} = T_p\int d^{p+1}x\, \frac{1}{(p+1)!}\epsilon^{\alpha_1\cdots \alpha_{p+1}}C_{\mu_1 \cdots \mu_{p+1}}\partial_{\alpha_1} X^{\mu_1}\cdots \partial_{\alpha_{p+1}} X^{\mu_{p+1}}.
\end{equation}
To cancel the trivial configuration energy, we only need a $C_{p+1}$
that has support along the directions parallel to the brane, but we
will allow more general coordinate dependence, and take into account
the gauge-field-dependent  terms in the WZ action as well.

Like the DBI action, the $C_{p+1}$ term is diffeomorphism invariant,
so one must fix the gauge before deforming the solution. Then the
pullback of the RR form potential is
\begin{equation}
S_{WZ}=T_p\int d^{p+1}x\, \sum_{n= 0}^p\frac{1}{(p-1+n)!}\epsilon^{\mu_1\cdots \mu_n \alpha_{1}\cdots\alpha_{p+1-n}} C_{\mu_1\cdots\mu_n a_1\cdots a_{p+1-n}}\partial_{\alpha_1}Y^{a_1}\cdots \partial_{\alpha_{p+1}}Y^{a_{p+1-n}}.
\end{equation}
Similar expressions can be found for terms involving gauge fields.
We will now study the lower dimensional D-branes individually. We
will assume throughout the calculation that $G_{ab}$ only depends on
the transverse coordinates $Y^a$.

We will use the following notation: $C^{(k)}$ denotes the RR form
potential of rank $k$. $\alpha'$ factors are absorbed in the gauge
fields $\tF_{MN}=2\pi\alpha' F_{MN}$, and we sometimes use
$B=\frac{1}{2}\epsilon^{ij}\tF_{ij}$ and $E_i=\tF_{0i}$ for magnetic
and electric fields.

\paragraph{D1 brane}

The Wess-Zumino action takes the simple form
\begin{align}
\notag S_{WZ,D1} &=T_1\int d^2 x\left[ \frac{1}{2}\epsilon^{\mu\nu} C^{(2)}_{\mu\nu}+2\frac{1}{2}\epsilon^{\mu\alpha}C^{(2)}_{\mu a}\partial_\alpha Y^a+\frac{1}{2}\epsilon^{\alpha\beta} C^{(2)}_{ab}\partial_\alpha Y^a \partial_\beta Y^b+\frac{1}{2}C^{(0)}\epsilon^{\mu\nu}\tF_{\mu\nu}\right]\\
&= T_1\int d^2x \left[
C^{(2)}_{01}+C^{(2)}_{0a}\d_1Y^a+C^{(0)}\tF_{01}\right],
\end{align}
Since we assume $\partial_0 Y^a=0$, the term proportional to
$\epsilon(\partial Y)^2$ vanishes.

The WZ contribution to the energy is
\begin{equation}
E_{WZ,D1}=-T_1\int dx \left[C^{(2)}_{01}+C^{(2)}_{0a}\d_1Y^a+2\pi\alpha' C^{(0)}F_{01} \right].
\end{equation}
The first term should be such that when added to the contribution from the DBI part the total energy is finite $T_1 C^{(2)}_{01}=\cE_{DBI}+\cdots$.

The contribution of the WZ terms to the stress tensor is
\begin{equation}
\Delta \Pi^1_{\ 1}=-T_1\left[ C^{(2)}_{0a}\d_1Y^a+2\pi\alpha' C^{(0)}F_{01}\right] +T_1 \left[C^{(2)}_{01}+C^{(2)}_{0a}\d_1Y^a+2\pi\alpha' C^{(0)}F_{01}\right]=T_1 C^{(2)}_{01}.
\end{equation}
So only the components of the $C_2$ RR potential along the
directions of the D1 modify the DC. The stress tensor has only one
component (we are assuming $G_{01}=0$):
\begin{equation}
\Pi^1_{\ 1}=\cE(g^{11}G_{ab}\partial_1 Y^a\partial_1 Y^b-1)+T_1 C^{(2)}_{01}.
\end{equation}
For a probe D1-brane in a background sourced by other parallel
D1-branes, $T_1 C^{(2)}_{01}=\cE$. If we also use that
$G_{ab}\partial_1 Y^a\partial_1 Y^b=g_{11}-G_{11}$, then
\begin{equation}
\Pi^1_{\ 1}=\cE(1-g^{11}G_{11}).
\end{equation}
Note that $g^{11}G_{11}=G_{11}/g_{11}\leq 1$, and $g^{11}G_{11}>0$. Therefore $\Pi^1_{\ 1}\geq 0$, being equal to zero only for the trivial configuration. For non-trivial configurations we would have
\begin{equation}
0 <\int dx\,\Pi^1_{\ 1}.
\end{equation}
In the absence of gauge fields there are no soliton solutions on the
D1-brane, in agreement with our previous analysis.

\paragraph{D2 brane}

For D2-branes the Wess-Zumino action is
\begin{align}
\notag S_{WZ,D2} &=T_2\int d^3 x\left[ C^{(3)}_{012}+\epsilon^{ij}C^{(3)}_{0ia}\partial_j
Y^a+\frac{1}{2}\epsilon^{ij}C^{(3)}_{0ab}\partial_i Y^a \partial_j
Y^b\right. \\
 &\left. +\frac{1}{3}C^{(1)}_0 B-\frac{1}{3}\epsilon^{ij} C^{(1)}_i E_j-\frac{1}{3}\epsilon^{ij}C^{(1)}_a\partial_i Y^a E_j\right],
\end{align}
From this expression we find the contribution to the energy
\begin{align}
\notag E_{WZ,D2} &=\int d^2x\,\cE_{WZ}=-T_2\int d^2 x \left[ C^{(3)}_{012}+\epsilon^{ij}C^{(3)}_{0ia}\partial_j
Y^a+\frac{1}{2}\epsilon^{ij}C^{(3)}_{0ab}\partial_i Y^a \partial_j
Y^b\right. \\
 &\left. +\frac{1}{3}C^{(1)}_0 B-\frac{1}{3}\epsilon^{ij} \left( C^{(1)}_i +C^{(1)}_a\partial_i Y^a \right)E_j\right].
\end{align}
Again, the first term should be such
 that when added to the contribution from the DBI part the total energy is finite $T_2 C^{(3)}_{012}=\cE_{DBI}+\cdots$.

The contribution of the WZ terms to the stress tensor is
\begin{align}
\notag \Delta \Pi^i_{\ j} &=-T_2\left[\epsilon^{ki}C^{(3)}_{0ka}\partial_j
Y^a+\epsilon^{ik}C^{(3)}_{0ab}\partial_j Y^a \partial_k
Y^b+\frac{1}{3}C^{(1)}_0 \epsilon^{ik}\tF_{jk}\right.\\
&\left.-\frac{1}{3}\epsilon^{ki} \left( C^{(1)}_k +C^{(1)}_a\partial_k Y^a \right)E_j-\frac{1}{3}\epsilon^{ik} \left(C^{(1)}_a\partial_j Y^a \right)E_k\right]-\delta^i_{\ j}\cE_{WZ}.
\end{align}

\paragraph{D3 brane}

For D3 branes the Wess-Zumino action is
\begin{align}
\notag S_{WZ,D3} &=T_3\int d^4 x\left[C^{(4)}_{0123}+\frac{1}{2}\epsilon^{ijk} C^{(4)}_{0ajk}\partial_i Y^a+\frac{1}{2}\epsilon^{ijk} C^{(4)}_{0abk}\partial_i Y^a\partial_j Y^b\right.\\
\notag &\left.+\frac{1}{3!}\epsilon^{ijk} C^{(4)}_{0abc}\partial_i Y^a\partial_j
Y^b\partial_k Y^c+\frac{1}{2\cdot
3!}\epsilon^{ijk}C^{(2)}_{0i}\tF_{jk}+\frac{1}{2\cdot
3!}\epsilon^{ijk}C^{(2)}_{0a}\partial_i Y^a \tF_{jk}\right.\\
 &\left.+ \frac{1}{2\cdot 3!}\epsilon^{ijk}C^{(2)}_{ij}\tF_{0k}+\frac{1}{3!}\epsilon^{ijk} C^{(2)}_{ia}\partial_j Y^a \tF_{0k}+\frac{1}{2\cdot 3!}\epsilon^{ijk}C^{(2)}_{ab}\partial_i Y^a\partial_j Y^b \tF_{0k}+\frac{1}{2\cdot 3!}C^{(0)}\epsilon^{ijk}\tF_{0i} \tF_{jk}
\right].
\end{align}
From this expression we find the contribution to the energy
\begin{align}
\notag E_{WZ,D3} &=-T_3\int d^3 x\left[C^{(4)}_{0123}+\frac{1}{2}\epsilon^{ijk} C^{(4)}_{0ajk}\partial_i Y^a+\frac{1}{2}\epsilon^{ijk} C^{(4)}_{0abk}\partial_i Y^a\partial_j Y^b\right.\\
\notag &\left.+\frac{1}{3!}\epsilon^{ijk} C^{(4)}_{0abc}\partial_i Y^a\partial_j
Y^b\partial_k Y^c+\frac{1}{2\cdot
3!}\epsilon^{ijk}C^{(2)}_{0i}\tF_{jk}+\frac{1}{2\cdot
3!}\epsilon^{ijk}C^{(2)}_{0a}\partial_i Y^a \tF_{jk}\right.\\
 &\left.+ \epsilon^{ijk}E_k\left( \frac{1}{2\cdot 3!}C^{(2)}_{ij}+\frac{1}{3!} C^{(2)}_{ia}\partial_j Y^a +\frac{1}{2\cdot 3!}C^{(2)}_{ab}\partial_i Y^a\partial_j Y^b +\frac{1}{2\cdot 3!}C^{(0)}\tF_{ij}\right)
\right].
\end{align}
The first term should be such that when added to the contribution from the DBI part the total energy is finite $T_3 C^{(4)}_{0123}=\cE_{DBI}+\cdots$.

The contribution of the WZ terms to the stress tensor is
\begin{align}
\notag \Delta \Pi^i_{\ j} &=-T_3\left[\frac{1}{2}\epsilon^{ikl} C^{(4)}_{0akl}\partial_j Y^a+\epsilon^{ikl} C^{(4)}_{0abl}\partial_j Y^a\partial_k Y^b
+\frac{1}{2}\epsilon^{ikl} C^{(4)}_{0abc}\partial_j Y^a\partial_k
Y^b\partial_l Y^c\right.\\
\notag &\left.+\frac{1}{
3!}\epsilon^{ikl}C^{(2)}_{0l}\tF_{jk}+\frac{1}{
3!}\epsilon^{ikl}C^{(2)}_{0a}\partial_l Y^a \tF_{jk}+\frac{1}{2\cdot
3!}\epsilon^{ikl}C^{(2)}_{0a}\partial_j Y^a \tF_{kl}\right.\\
 &+ \epsilon^{kli}E_j\left( \frac{1}{2\cdot 3!}C^{(2)}_{kl}+\frac{1}{3!} C^{(2)}_{la}\partial_l Y^a +\frac{1}{2\cdot 3!}C^{(2)}_{ab}\partial_l Y^a\partial_l Y^b +\frac{1}{2\cdot 3!}C^{(0)}\tF_{kl}\right)\\
&\left.+ \epsilon^{kil}E_l\left(\frac{1}{3!} C^{(2)}_{ka}\partial_j Y^a + \frac{1}{ 3!}C^{(2)}_{ab}\partial_k Y^a\partial_j Y^b -\frac{1}{ 3!}C^{(0)}\tF_{jk}\right)
\right] -\delta^i_{\ j}\cE_{WZ}.
\end{align}

\subsubsection{Examples in flat space}

\paragraph{D1 brane with electric field}

We consider a D1 brane in some background, described by a DBI
action and Chern-Simons action. The vanishing of the stress tensor imposes the condition
\begin{align}
0=&
2C^{(2)}_{01}-\frac{e^{-\phi}G_{00}G_{11}}{(-G_{00}g_{11}-\tF_{01}^2)^{3/2}}\left(2\tF_{01}^2+G_{00}g_{11}
\right)~.
\end{align}
Strictly speaking we need to impose only that the integral of the stress tensor vanishes, if this condition can be satisfied there could be BPS solitons.  If we allow electric fields $\tE=\tF_{01}$
and $Y'$ but remain in a flat background we have
\begin{align}\label{eq:Derrickflat}
0= 2C_{01} - e^{-\phi}\frac{1+Y'^2-2\tE^2}{(1+Y'^2-\tE^2)^{3/2}}~.
\end{align}
We know (already from the action, in fact) that we must have
$\tE^2<1+Y'^2$. Let's assume no backreaction of the probe on the
dilaton or the background, and take $2C_{01}=e^{-\phi}$~. This
implies
\begin{align}
0=\ 2C_{01}\left\{ 1 - \frac{1+Y'^2-2\tE^2}{(1+Y'^2-\tE^2)^{3/2}}
\right\}~.
\end{align}
The solution must also satisfy equations of motion. Since these have
no direct dependence on $x^1$, we can write them in terms of
integrals of motion,
\begin{align}
\frac{Y'}{\sqrt{1+Y'^2-\tE^2}}=&e^\phi Q_Y\\
-\frac{\tE}{\sqrt{1+Y'^2-\tE^2}}=&e^\phi Q_F\\
\end{align}
where $Q_Y$ and $Q_F$ are constants. Adding and subtracting these we
have
\begin{align}
\frac{(Y'-\tE)}{\sqrt{1+Y'^2-\tE^2}}&=e^\phi (Q_Y+Q_F)\qquad\qquad \frac{(Y'+\tE)}{\sqrt{1+Y'^2-\tE^2}}=e^\phi (Q_Y-Q_F) \nonumber\\
\Rightarrow& \frac{(Y'^2-\tE^2)}{1+Y'^2-\tE^2}=e^{2\phi}
(Q_Y^2-Q_F^2)\equiv Q^2(x)
\end{align}
where in the second line we multiplied the two equations together.
We can simplify this to give
\begin{align}
Y'^2-\tE^2=\frac{Q^2}{1-Q^2}~.
\end{align}
Using this result we can simplify the Derrick-type condition of
equation (\ref{eq:Derrickflat}):
\begin{align}
0= e^{-\phi}\left\{ 1-\sqrt{1-Q^2}\left[
1-(1-Q^2)\tE^2\right]\right\}~.
\end{align}
This gives
\begin{align}
\tE^2&=-(1-Q^2)^{-3/2}\left( 1-\sqrt{1-Q^2}\right)
\end{align}
This would require $\tE^2<0$: a contradiction. We can now see that
on the D1-branes, there are no BPS solitons.

\paragraph{D2 brane with magnetic field}
In \S \ref{bpsbrane} we considered BPS Abelian vortices. We will see
now that when a magnetic field is turned on in the D-brane action,
such configurations cannot exist. We restrict to $E_i=0$,  we also
require $C^{(3)}_{012}=e^{-\phi}$ in order to have finite energy
configurations,\footnote{We are assuming that magnetic fields vanish
at infinity, otherwise the energy would diverge with this
subtraction.} and take the same duality condition for scalar fields
as before \eqref{vortexBPS}. The induced metric and DBI energy
density are
\begin{equation}
g_{ij}=(1+(\partial Y)^2)\delta_{ij}, \ \ \cE_{DBI}=\sqrt{(1+(\partial Y)^2)^2+B^2}.
\end{equation}
We will also need
\begin{equation}
\det M=\det \hat M=1+\frac{B^2}{(1+(\partial Y)^2)^2}.
\end{equation}
The contribution from the magnetic fields to the stress tensor \eqref{stressDBI} is
\begin{equation}
{\Pi_B}^i_{\ j}=\frac{1}{2}\frac{\cL_{DBI}}{\det\hat{M}}\frac{\delta \det \hat M}{\delta F_{nl}}\left((F_n^{\ i}\delta_l^k+F_n^{\ k}\delta_l^i)G_{ab}\partial_k Y^a\partial_j Y^b-2\delta^i_n F_{jl} \right)=\frac{\delta^i_{\ j}B^2}{(1+(\partial Y)^2)^2\sqrt{\det\hat{M}}}.
\end{equation}
Here we have used $\cL_{DBI}=-\cE_{DBI}$, the expression for the
induced metric, $F_{ij}=B\epsilon_{ij}$ and
\begin{equation}
\frac{\delta \det \hat M}{\delta F_{nl}}=\frac{B\epsilon^{nl}}{(1+(\partial Y)^2)^2}.
\end{equation}
One is left with
\begin{equation}
\Pi^i_{ \ j}=\delta^i_{\ j}\frac{\sqrt{\det\hat{M}}-1}{\sqrt{\det\hat{M}}}.
\end{equation}
For $B=0$ (usual BPS vortices) this is exactly zero, but otherwise
$\Pi^i_{ \ j}\geq 0$, so its integral cannot vanish and there are no
vortex solutions satisfying \eqref{vortexBPS} for $B\neq 0$.

\section{Some applications in string theory}

\subsection{ Flavor branes in M theory}

Another example of the application of DC's to a system of flavor
branes is that of the holographic QCD model discussed in
\cite{Aharony:2010mi}. Consider embedding in a background of $N$
D4-branes, a probe brane system that includes $p$ D4-branes
stretched between two perpendicular NS5-branes (see
figure~\ref{branesetup}). Viewing the $N$ D4-branes as M5-branes
wrapped around the M-theory circle, their eleven dimensional
geometry is given by

\FIGURE[t]{
\includegraphics[width=10cm]{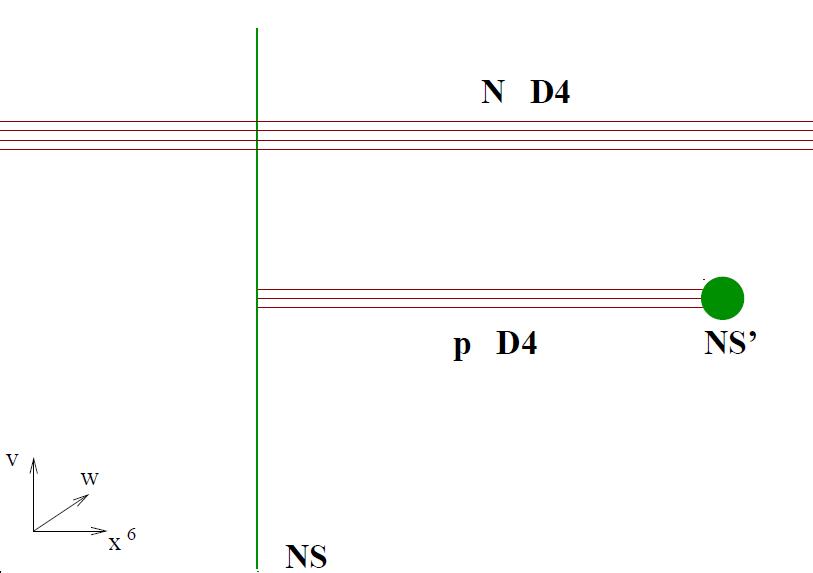}
\caption{The brane system.\label{branesetup}}
}

\bea\label{TheMetric}
ds^2&=&H^{-1/3}\left(dx_\mu^2+dx_6^2+dx_{11}^2\right)+
H^{2/3}\left(|dv|^2+|dw|^2+dx_7^2\right),\nonumber\\
C_6&=&H^{-1}d^4x\wedge dx_6\wedge dx_{11},\qquad
H=1+\frac{\pi\lambda_N l_s^2}{|\vec r-\vec r_0|^3}\,, \eea where
$\lambda_N=g_{YM}^2 N$ is the $4+1$ dimensional `t Hooft coupling.
$\mu=0,1,2,3$, $\vec r=(v,w, x^7)$ labels position in $\IR^5$, and
$\vec r_0\in \IR^5$ is the position of the $N$ D4-branes in
figure~\ref{branesetup}.

The shape of the curved fivebrane we are interested in is obtained
by plugging in the ansatz \bea\label{nsansatz}
v=u(x_6)e^{i\phi(x_{11})}\sin \a(x_6),\qquad
w=u(x_6)e^{-i\phi(x_{11})}\cos \a(x_6), \eea into the  M5-brane
worldvolume action. The induced metric on the fivebrane
corresponding to (\ref{nsansatz}) is \bea
ds_{ind}^2=H^{-1/3}\left\{dx_\mu^2+\left[1+H\left((u\a')^2+(u')^2\right)\right]dx_6^2+
\left(1+H(u{\dot \phi})^2\right)dx_{11}^2\right\}, \eea where
$u'\equiv \pa_{x_6} u$, $\a' \equiv \pa_{x_6} \a$ and
$\dot\phi=\pa_{x_{11}}\phi$. The Lagrangian is \bea\label{Laguncomp}
L=-H^{-1}\sqrt{1+H(u{\dot
\phi})^2}\sqrt{1+H((u\a')^2+(u')^2)}+H^{-1}. \eea The equations of
motion imply that ${\dot\phi}$ must be constant; thus
$\phi=x_{11}/pR=x_{11}/\lambda_p$. The Noether charges $J, P_6$
associated with the invariance under the shift of $\alpha$ and
$x_6$, respectively, are then given by \bea\label{ccc} J
&=&\frac{u^2\a'\sqrt{1+ H
u^2/\lambda_p^2}}{\sqrt{1+H\left[(u\a')^2+(u')^2\right]}},\CR
 P_6&=& H^{-1}-\frac{H^{-1}\sqrt{1+H u^2/\lambda_p^2}}{\sqrt{1+H\left[(u\a')^2+(u')^2\right]}}.
\eea

The profile of the probe M5 brane is described by the $v$ and $w$,
which are expressed in terms of $u$ and $\a$ given in
(\ref{nsansatz}). The latter are determined by the effective 1+1
dimensional action $S=\int dt dx_6 L$. The corresponding energy
for a static configuration is $E=-S$.

Applying a uniform scaling deformation, and minimizing $E(\lambda)$,
we get \be
\left.\frac{dE(\lambda)}{d\lambda}\right|_{\lambda=1}=\int dx\
H^{-1}\left\{ 1 -\frac{\sqrt{1+H
u^2/\lambda_p^2}}{\sqrt{1+H\left[(u\a')^2+(u')^2\right]}}\right\}=0
\ee The integrand is in fact the Noether charge density associated
with the invariance under the shift $x_6\rightarrow x_6 + a$ where
$a$ is a constant. Stated differently, had we taken the action to be
$0+1$ dimensional with $x_6$ as the time direction, the integrand
would have been the energy density -- as it is termed in
\cite{Aharony:2010mi}. The solution for the profile in
\cite{Aharony:2010mi}  was taken to be the so-called supersymmetric
profile where this ``energy density" vanishes.  The corresponding
profile takes the form \bea\label{NSUsol}
u=\sqrt{\frac{J\lambda_p}{2}}\sqrt{2\cosh
\frac{2x_6}{\lambda_p}},\qquad
\tan\a=\exp\left(2x_6/\lambda_p\right), \eea which fulfills a
Derrick-type condition. Requiring that the ``energy density"
associated with the shift invariance  of a space coordinate should
vanish, is equivalent to the Hamiltonian constraint. This occurs
also in the applications to gravitational backgrounds discussed
below. It is also easy to see that this holds in a more general
setup. The reason Derrick's condition coincides with the zero
``energy'' condition is that an infinitesimal dilatation is
equivalent to an infinitesimal translation with a space-dependent
parameter, so  the term appearing in the variation of the energy is
proportional to the generator of translations along $x$. Consider
the following 1+1 dimensional Lagrangian density  for $N$ degrees of
freedom \be {\cal L}= -\sqrt{g_0(x) - \sum_{i=1}^N g_i [
(\partial_t\phi_i)^2- (\phi_i')^2 ]}- V(\phi_i) \ee For a static
configuration, the energy density is ${\cal H}=-{\cal L}$ and
Derrick's condition reads \bea
\left.\frac{dE(\lambda)}{d\lambda}\right|_{\lambda=1}&=&\int dx\
\left\{ -\sqrt{g_0(\phi_i) + \sum_{i=1}^N g_i [ (\phi_i')^2 ]}-
V(\phi_i)+ \frac{\sum_{i=1}^N g_i [ (\phi_i')^2 ]}{\sqrt{g_0(x)
+\sum_{i=1}^N g_i [ ( \phi_i')^2 ]}}\right \}\CR
 &=& -\int dx\ \left\{ V(\phi_i) + \frac{g_0(\phi_i) }{\sqrt{g_0(x)  +\sum_{i=1}^N g_i [ ( \phi_i')^2 ]}}\right \} =0 ~.\CR
\eea The expression in brackets is identical to the ``energy" of the
$0+1$  action where  we have taken $x$ to be ``time."

\subsection{Application to gravitational backgrounds}

The DC on D-brane soliton solutions of super-gravity (SUGRA) actions
are also interesting. The bosonic part of the
 SUGRA action in $D$ dimensions takes the form
\bea       \label{eq:TheAction}
S &=& \int d^{D} x \sqrt{G} e^{-2 \phi}
               \left( R+ 4(\pa\phi)^2 + \frac{c}{\alpha^\prime} \right)\CR
 && - \frac{e^{-2 \phi}}{2}  \int H_{(3)} \wedge \star H_{(3)}
    - \sum_{p} \frac12   \int F_{(p+2)} \wedge \star F_{(p+2)} ,
\eea
where $\phi$ is the dilaton,
 $F_{p+2}$ is a RR form
that corresponds to a Dp-brane with $n=p+1$ dimensional world
volume, $ H_{(3)}$ is the NS three form and $
\frac{c}{\alpha^\prime}=\frac{10-D}{\alpha^\prime} $ is the
non-criticality central charge term.

Let's assume now that the  metric in the \emph{string} frame   depends only on  the
radial coordinate $\tau$.
It takes the form
\begin{equation}       \label{stringmetric}
l_s^{-2}ds^2 = d\tau^2 +  e^{2\lambda(\tau)} dx_{\|}^2 +e^{2\nu(\tau)} d\Omega_k^2
\end{equation}
where $D= n+k+1$,  $dx^2_{\|}$ is $n$ dimensional flat metric, and
$d\Omega_k^2$ is a $k$ dimensional sphere.

Upon substituting the metric (\ref{stringmetric}) into  the action and performing
the integration  one finds\cite{Kuperstein:2004yk}
\bea \label{actionrho}
S &=& l_s^{-2}\int d\rho  \left( \left[-n(\lambda')^2 -k(\nu')^2 +(\varphi')^2
+ c e^{-2 \varphi} + (k-1)k e^{-2\nu-2 \varphi} \right]\right)+ S_{RR} + S_{NS}\CR
\notag S &=& l_s^{-2}\int d\rho  \left( \left[-n(\lambda')^2 -k(\nu')^2 +(\varphi')^2
+ c e^{-2 \varphi} + (k-1)k e^{-2\nu-2 \varphi} \right]\right.\\
& & \left.  - Q_{RR}^2 \rho e^{n \lambda -k \nu - \varphi}- Q_{NS}^2
e^{- 2k \nu- 2 \varphi}\right) \eea where $d\tau=
-e^{-\varphi}d\rho$, $(A)'=\pa_\rho A$ and $
\varphi=2\phi-n\lambda-k\nu $ and the NS term is relevant only for
$k=3$.

The action (\ref{actionrho})   can be viewed as a 1+1 dimensional
action for three static  scalar fields $\lambda, \nu$ and $\varphi$,
subject to the potential \be V = - (c + (k-1)k e^{-2\nu})
e^{-2\varphi}+{Q_{RR}^2} e^{n\lambda-k\nu-\varphi} +Q_{NS}^2 e^{- 2k
\nu- 2 \varphi}.
\end{equation}
In fact the dimensionally reduced system is characterized not only by the action (\ref{actionrho}) but also by
 the so called ``Hamiltonian constraint", which is the Gauss' law associated with fixing
   $g_{\tau\tau}=1 $
 and takes the following form
\be
{\cal H}=n(\pa_\tau\lambda)^2 +k(\pa_\tau\nu)^2 -(\pa_\tau\varphi)^2
+ c  + (k-1)k e^{-2\nu}- Q_{RR}^2 e^{n\lambda-k\nu+\varphi}-
Q_{NS}^2 e^{-2k \nu } =0.
\end{equation}
Treating the system as that of 1+1 dimensional system of  three scalar fields $\lambda,\nu,\varphi$ subjected to a potential  , we are led to Derrick's condition (\ref{derthe}) that takes the form
\bea\label{derthegrav}
\notag \left.\frac{dE(\lambda)}{d\lambda}\right|_{\lambda=1} &=&   \int d \tau \left [ n(\pa_\tau\lambda)^2 +k(\pa_\tau\nu)^2 -(\pa_\tau\varphi)^2
+ c  + (k-1)k e^{-2\nu}\right.\\
& & \left. - Q_{RR}^2 e^{n\lambda-k\nu+\varphi}-
Q_{NS}^2 e^{-2k \nu }\right ]
\eea

It is now straightforward to realize that the integrand of this integral is precisely the Hamiltonian constraint ${\cal H}=0$ and hence by construction static  solutions of the system (\ref{actionrho}) is in accordance with Derrick's theorem. These  include the D-brane solitons of the critical theory as well as the D-brane solutions of non-critical dimensions\cite{Kuperstein:2004yk}.

\subsection{Wilson lines}

We noted above that finite-size spaces involve some subtleties in
the DC language. We now describe how to study Wilson lines of finite
length in this scheme, that we assume can be described using the
Nambu-Goto action either in flat space or in a holographic setup. In
principle the analysis can also be extended to probe branes with
boundaries. In these cases, equation \eqref{manton} may be modified
by boundary terms.

Using the conservation of the energy-momentum tensor, $\d_i
\Pi^i_{\phantom{i}j}=0$ for some vector $\xi^j$,
\begin{align}
\int_{\mathcal{M}}d^dx \ \d_i
(\xi^j\Pi^i_{\phantom{i}j})=\int_{\d{\mathcal M}} d^{d-1}x \ \xi^j
\hat{n}_i\Pi^i_{\phantom{j}j}~,
\end{align}
where $\hat{n}$ is the normal to the spatial boundary of the
manifold $\mathcal{M}$. Using $\xi^j = \lambda^j_{\phantom{j}k}x^k$ for
arbitrary constant $\lambda^j_{\phantom{j}k}x^k$, this gives the conditions
\begin{align}
\int_{\mathcal{M}}d^dx \ \Pi^i_{\phantom{i}j}=\int_{\d{\mathcal M}}
d^{d-1}x \ x^i\hat{n}_k\Pi^k_{\phantom{j}j}~.
\end{align}
The above expression holds for bounded spaces of arbitrary
dimension. For the simple case of a Wilson line, there is only a
single component in the stress tensor $\Pi^1_{\ 1}\equiv \Pi$. If
the background metric does not depend explicitly on $x$, then the
conservation (and lack of time dependence) for $\Pi$ implies that it
is constant along the spatial direction $\partial_x \Pi=0$. We will
assume that the coordinate $x$ is restricted to the interval $x\in
[x_1,x_2]$, and that there is a soliton solution with profile
$Y^a(x)$, satisfying some fixed boundary conditions at the endpoints
of the interval
\begin{equation}
F_n\left[Y^a(x_n),\partial_x Y^a(x_n)\right]=0, \ \ n=1,2.
\end{equation}
We now consider the deformed configuration $\tilde Y^a(x)=Y^a(\lambda x)$, in the interval with boundary at $\tilde x_n=x_n/\lambda$.
Clearly, the new configuration will satisfy the following  boundary conditions
\begin{equation}
F_n\left[\tilde Y^a(\tilde x_n),\frac{1}{\lambda}\partial_x \tilde Y^a(\tilde x_n)\right]=0, \ \ n=1,2.
\end{equation}
For both Dirichlet $F_n[Y,\partial Y]=Y-Y_n$ and Neumann
$F_n[Y,\partial Y]=\partial Y$, the form of the boundary conditions
is preserved. In the first case, the position of the endpoints in
the `transverse' directions remains fixed. Let us focus on this case
in what follows.

It is convenient to write the energy of the soliton solution as an
integral over the full real line
\begin{equation}
E[Y^a]=\int_{-\infty}^\infty d\sigma \cE[Y^a,\partial_\sigma Y^a] G[\sigma],
\end{equation}
where $\cE$ is the energy density and
\begin{equation}
G[\sigma]=\Theta(\sigma-x_1)-\Theta(x_2-\sigma).
\end{equation}
The deformed configuration preserving the Dirichlet boundary condition is $\tilde Y^a(\sigma)=Y^a(\lambda\sigma)$ and $\tilde G(\sigma)=G(\lambda\sigma)$. The energy is
\begin{equation}
E[\tilde Y^a]=\int_{-\infty}^\infty d\sigma \cE[\tilde Y^a,\partial_\sigma \tilde Y^a] \tilde G(\sigma).
\end{equation}
We now rescale the spatial coordinate $\sigma\to \sigma/\lambda$, then
\begin{equation}
E[\tilde Y^a]=\int_{-\infty}^\infty d\sigma \frac{1}{\lambda}\cE[Y^a,\lambda \partial_\sigma Y^a]G(\sigma)=\int_{x_1}^{x_2} dx \frac{1}{\lambda}\cE[Y^a,\lambda \partial_x Y^a].
\end{equation}
This leads to the usual result where the variation of the energy is proportional to the integral of the stress tensor
\begin{align}
\delta E \sim \int_{x_1}^{x_2}dx \ \Pi=(x_2-x_1)\Pi\equiv \Delta
x\Pi~,
\end{align}
where we have used the fact that $\Pi$ is constant. The variation is
thus proportional to the length of the interval. This happens even
for the ground state configuration because we are allowing the size
of the interval to vary. For instance, if we think about the
holographic calculation of a quark-antiquark potential, the deformed
configuration still corresponds to a string ending at the AdS
boundary, but the position of the endpoints along the spatial
directions of the field theory is allowed to change. Clearly the
energy is extremized when the two endpoints coincide and the string
shrinks to zero size. This does not reflect a dynamical instability
because the conserved quantity $\Pi$ is different for different
sizes of the interval.

One should keep in mind that for the interesting case of Wilson lines in AdS backgrounds (as discussed at length in \cite{Kol:2010fq}), the bare
energy derived from the action is not finite and must be
renormalized. A Wilson line with its two endpoints on the AdS
boundary will tend to dip down into the space. If one simply uses a
subtraction scheme in which one subtracts off the energy, $E_{sub}$,
of two straight Wilson lines to yield a finite energy, the analysis
is completely unaffected.

An interesting extension of this analysis will be to introduce a
deformed configuration with fixed boundary conditions in the same
interval. However, this is not possible for a Dirichlet condition
and the simple re-scaling presented here, where $\lambda$ is
constant, since this would require that $\lambda(x_n)=1$ at the
endpoints of the interval.

\section{Summary and open questions}

The equations of motion of any interacting physical system are
generically non-linear. There are only few cases for which exact
solutions to these equations are known. It is thus interesting and
important to derive constraints on physically realized solutions to
these equations, as we have done here.

We have developed, analyzed and applied what we refer to as
``deformation constraints" (DC). A special case of these
constraints, used to prove Derrick's theorem
\cite{Derrick:1964ww,Hobart}, has been known for fifty years.
Another class of constraints were proposed recently by Manton
\cite{Manton:2008ca}. We rederived these two types of constraints by
considering generalized deformations of energy-minimizing soliton
solutions, including also internal symmetries. We have demonstrated
the use of the DC in several soliton solutions like the magnetic
monopole and the abelian Higgs model, and have obtained some novel
results in the analysis of D-brane system solitons. We have applied
the DC to a variety of D-brane actions that include the DBI as well
as  WZ terms. We have also discussed the application to flavor
branes in M theory, Wilson lines and general static solutions of
gravitational backgrounds. Other special cases that we have
considered are non-linear generalizations of electromagnetism (of
which the DBI action is a special case).

 We have derived the following concrete constraints from the DC applied to branes:
\begin{itemize}
\item
There can be no solitons of the DBI action (for arbitrary Dp-branes)
in which the transverse directions depend only on one worldvolume
coordinate.
\item
While solitons are excluded for Maxwell's theory in four dimensions
(4D), there is no such exclusion of solitons for DBI
electromagnetism in 4D. Other non-linear generalizations may even
admit magnetic solitons.
\item
For  $D1$ branes with electric field in flat space-time, there are
no BPS solutions living on the brane worldvolume.
\item
For $D2$ branes in flat space time, Abelian BPS vortices completely
disappear if a magnetic field is turned on on the $D2$ brane.
\end{itemize}

The DC may serve as a  useful tool for assessing classical static
solutions in the context of ``ordinary"  field theories,  D-brane
models and gravity and string actions. The application of  DC to
other areas, and their extension to infinite-energy configurations,
remain to be explored. We list some of these interesting directions
here:
\begin{itemize}

\item
We have considered solitons in field theory and D-brane models.
There are a variety of other systems, however, in which these DC may
find fruitful application.

Solitonic configurations appear in optical\cite{opticalsoliton} and
magnetic systems, hydrodynamics and even in the physics of proteins
and DNA. Soliton solutions exist also for equations of motion
associated with non-relativistic actions which are not Lorentz
invariant. As the DC represent a general framework for constraining
finite energy static solutions, they make prove useful for
identifying solutions of systems governed by the KDV and non-linear
Schr\"{o}edinger equations. One could also use the constraints to
build new Lagrangians that admit (or exclude) solitons by design.

It may also be interesting to continue the work of section \S  6,
using the DC to search for gravitational solutions in cases where
the kinetic terms and the ``potential" are positive definite (which
is not true of all gravitational systems).
Other interesting extensions may include BIon-like solutions
\cite{Gibbons:1997xz, Callan:1997kz} which are not fully smooth and
have a singularity and an associated charge.

The DC technique may even facilitate the solution of non-linear
differential equations which are not necessarily related to physical
systems. Given an equation, assume one can construct a Lagrangian
density for which it is an equation of motion.  One can then
incorporate the time direction by adding  a kinetic term  to the
Lagrangian density. The minimization of the corresponding energy
with respect to deformations of the solutions will constitute a DC
for this non-differential equation.

\item
Classical solutions of scalar field theories frequently appear in
cosmology. In that case, one is not interested in static solutions
but in solutions which are only time dependent. It may be possible
to develop constraints similar to the DC for such configurations.

\item
Another related question is whether one can impose similar
conditions by deforming solutions and minimizing energy for
quantized solitons\cite{Dashen:1974cj}.  The main difference is that
the classical energy can receive corrections from quantum
fluctuations around the semiclassical soliton configuration. This
configuration is in general different from the classical solution
and requires solving the equation derived from the effective action
in a self-consistent way.

\item
It was recently discovered\cite{Gauntlett:1997ss,Nakamura:2009tf}
that certain holographic models (both top-down and bottom-up) have
ground states which are spatially modulated -- that is,
translational invariance is spontaneously broken. In these cases the
energy is not finite. One must therefore develop new tools to apply
the deformation constraints locally\cite{DHS}.

 \item
Perhaps the best-known solitons are those associated with 1+1
dimensional models, such as the sine-Gordon model. This model is
integrable. As a consequence, it -- and others like it -- admit an
infinite set of conserved currents and associated charges. It would
be interesting to generalize the DC to derive constraints on the
spatial components of the infinite tower of conserved currents. It
is not obvious a priori whether these higher order constraints will
be independent, or whether they will follow automatically from the
leading order constraints.
\item
The DC presented in section \S 2 include all possible spatial
transformations that are linear in the coordinates. Special
conformal transformations, which accompany scale transformations of
any conformal invariant system, meanwhile, are quadratic in the
coordinates at infinitesimal order.
The quadratic deformations of the special conformal transformations
can be generalized to any quadratic transformation, or as
deformations at any order in the coordinates. These may imply the
vanishing of integrals of the stress tensor multiplied by higher
powers of the coordinates, implying a more rapid decay of the stress
tensor at large radial coordinates.

\item

In section \S 2 we touched briefly on the relation between BPS
configurations and the vanishing of the stress tensor.  In these
cases the stress tensor (not its integral) vanishes for all
components.
The interplay between the BPS condition, self-duality of solutions,
and the vanishing of the stress tensor deserve further
investigation, as do similar relations for global currents.

For instance, BPS solitons are commonly defined as solutions that
saturate the energy-charge bound. In supersymmetric theories this
implies that they conserve half of the supersymmetries. One can then
use the supersymmetric Ward identity to show that the components of
the stress tensor vanish. However, one could define the soliton from
the stress tensor, show that it must preserve half of the
supersymmetries, and then recover the relation between the charge
and the energy.

In all of the  examples we consider, BPS solitons break half of the
supersymmetries. It would be interesting to consider whether the
same conditions on the stress tensor are satisfied for solitons that
break a larger number of supersymmetries.

\end{itemize}

\section*{Acknowledgements}

C.H. would like to thank Eduardo Guendelman for useful
conversations. The work of  C.H. and J.S. is partially supported by
the Israel Science Foundation (grant 1665/10).


\appendix

\section{Examples of deformation constraints  on branes}
\label{app:derrickbrane}

\subsection{Probe branes in brane backgrounds }\label{sec:probebranes}

Particular examples where the brane action is finite for infinitely
extended configurations happen when the DBI action is supplemented
by a contribution from the RR fluxes that cancels out for trivial
configurations. This is the case when we have a Dp-brane background,
and compute the action for a Dp-probe in the background (extended
along all of the same directions as the background branes). The
near-horizon metric for the Dp background takes the form
\begin{equation}
ds^2=\left(\frac{R}{r}\right)^{(7-p)/2}dr^2+\left(\frac{r}{R}\right)^{(7-p)/2}\eta_{\mu\nu}dx^\mu
dx^\nu+R^2\left(\frac{r}{R}\right)^{(p-3)/2}d\Omega_{8-p}^2.
\end{equation}
Here $d\Omega_{8-p}^2$ is the metric of a unit $(8-p)$-sphere,
$S^{8-p}$. The ``conformal boundary'' is at $r\to \infty$. The
dilaton has the form
\begin{equation}
e^\phi=e^{\phi_0}\left(\frac{R}{r}\right)^{(7-p)(3-p)/4}.
\end{equation}
The potential  for $F_{p+2}=d C_{p+1}$ takes the form
\begin{equation}
C_{p+1}=\frac{1}{g_s}\left(\frac{r}{R} \right)^{7-p} dx^0\wedge
\cdots \wedge dx^p.
\end{equation}
The action  on the probe $Dp$-branes, adding both DBI and WZ terms
is
\begin{align}
S&=-T_{p}\int d^{p+1}x \left[e^{-\phi}\sqrt{-g} -C_{p+1}\right] =
-\frac{T_p}{g_s}\int d^{p+1}x\left[ \sqrt{-g}-
\left(\frac{r}{R}\right)^{7-p}\right].
\end{align}
We will restrict now to the simplest case where we assume that the
soliton is determined by a function $r(x^1)$ describing the change
in the radial position of the brane along a single spatial
direction. The embedding is
\begin{equation}
X^\mu=x^\mu, \ \ Y^{p+1}=r(x^1), \ \ Y^{a>p+1}=0.
\end{equation}
The action becomes (with $R=1$ for simplicity)
\begin{equation}
S=-\int d^{p+1}x L=-\int d^{p+1}x
r^{7-p}\left[\sqrt{1+r^{p-7}{r'}^2}-1\right].
\end{equation}
The energy density is simply minus the Lagrangian density
\begin{equation}
E=\int d^p x r^{7-p}\left[\sqrt{1+r^{p-7}{r'}^2}-1\right].
\end{equation}
We now follow Derrick's procedure and evaluate the energy for the
rescaled configuration $r(\lambda x^1)$, where we assume that $r(x)$
is a solution to the equations of motion. After we change
coordinates $x^1\to x^1/\lambda$ the energy becomes
\begin{equation}
E(\lambda)=\int d^p x \lambda^{-p} r^{7-p}\left[\sqrt{1+\lambda^2
r^{p-7}{r'}^2}-1\right].
\end{equation}
Then, Derrick's condition is
\begin{equation}
0=\left.\frac{dE}{d\lambda}\right|_{\lambda=1}=\int d^p x\left(
-pL+\frac{{r'}^2}{\sqrt{1+r^{p-7}{r'}^2}}\right).
\end{equation}
Note that for $r'=0$ we have that $L=0$, so this is satisfied for
the trivial solution, but as we saw in the general analysis soliton
solutions are not discarded in principle. However, we will show that
they are actually not allowed by showing that if Derrick's condition
was satisfied, the solution would not be a minimum of the energy,
but rather a maximum. First note that if there is a solitonic
solution it must be true that
\begin{equation}\label{eq:L}
L=\frac{1}{p}\frac{{r'}^2}{\sqrt{1+r^{p-7}{r'}^2}}.
\end{equation}
We now take the second derivative of the energy
\begin{equation}
\left.\frac{d^2 E}{d\lambda^2}\right|_{\lambda=1}=\int d^p x\left(
p(p+1)L-(2p-1)\frac{{r'}^2}{\sqrt{1+r^{p-7}{r'}^2}}-\frac{{r'}^4}{(1+r^{p-7}{r'}^2)^{3/2}}\right).
\end{equation}
Solving for $L$ using \eqref{eq:L} we have that
\begin{equation}
\left.\frac{d^2 E}{d\lambda^2}\right|_{\lambda=1}=\int d^p
x\left(-(p-2)\frac{{r'}^2}{\sqrt{1+r^{p-7}{r'}^2}}-\frac{r^{p-7}{r'}^4}{(1+r^{p-7}{r'}^2)^{3/2}}\right).
\end{equation}
This is negative for $p\geq 2$, which means there are no solitonic
configurations of this kind. For $p=1$
\begin{equation}
\left.\frac{d^2 E}{d\lambda^2}\right|_{\lambda=1}=\int d
x\frac{{r'}^2}{\sqrt{1+r^{-6}{r'}^2}}\left(1-\frac{r^{-6}{r'}^2}{1+r^{-6}{r'}^2}\right)=\int
d x\frac{{r'}^2}{(1+r^{-6}{r'}^2)^{3/2}}.
\end{equation}
which is manifestly positive, so solitons are in principle allowed
for D1 branes even taking into account this more restrictive
condition. A question is whether the extremality condition can
actually be satisfied once we use the equations of motion. The
answer is yes, but for the D1 brane
\begin{equation}
\left.\frac{dE}{d\lambda}\right|_{\lambda=1}\sim
-L+\frac{{r'}^2}{\sqrt{1+r^{-6}{r'}^2}}=\frac{\delta L}{\delta
r'}r'-L \equiv \pi^x,
\end{equation}
which is just a constant of motion for the D1 brane. The extremality
condition requires $\pi^x=0$. However this gives as a solution
$r'=0$, so there are actually no solitons.

Although we have ruled out only a very restricted class of
configurations, this shows explicitly that Derrick's condition can
be naturally extended to impose much stronger constraints.

\subsection{D3 brane with electric and magnetic fields}

Before imposing Derrick's conditions on the DBI action  recall the
conditions on the ordinary Maxwell theory. The energy is \be E=
\frac12\int d^d x [ (\vec B)^2 + (\vec E)^2]. \ee The scaling of the
electric and magnetic field as above, namely,  $ \vec E\rightarrow
\lambda \vec E$ and $ \vec B\rightarrow \lambda^2 \vec B$. Thus the
scaled energy is \be E(\lambda) = \frac12\int d^dx \lambda^{-d} [
\lambda^4(\vec B)^2 + \lambda^2(\vec E)^2]. \ee
 Derrick's condition
is therefore \be\left.\frac{dE(\lambda)}{d\lambda}\right|_{\lambda=1}=
\frac12\int d^d x  [(4-d) (\vec B)^2 -(d-2) (\vec E)^2]=0. \ee For
$d=3$ spatial dimensions (anti) self-dual configurations with $\vec
E =\pm\vec B$ obviously fulfill Derrick's condition but as usual the
actual condition is that the integral and not necessarily the
integrand will vanish.

We assume as background fields $C_{012}=e^{-\phi}=1$ in order to
have finite energy configurations. For simplicity we will assume
that the scalar fields are trivial. The corresponding action in flat
space takes the form \be S_{DBI}= T_3\int d^4 x \left[ 1-\sqrt{ 1-
(\vec E)^2 + (\vec B)^2 - (\vec E\cdot \vec B)^2 }\right] . \ee The
energy density associated with this action reads \be {\cal E}=
\sqrt{ 1+ (\vec B)^2 + (\vec D)^2 + (\vec B\times  \vec D)^2 } -1,
\ee
 where $\vec D$ is given by $\frac{\pa{\cal L}}{\pa\vec E}$.
 Consider first the case of only magnetic fields turned on. In this case
 \be\label{HBlambda}
 {\cal L}=1-\sqrt{1+ (\vec B)^2} \qquad  {\cal E}= \sqrt{1+ (\vec B)^2}-1
 \ee
 The scaled energy is
 \be
 E_B= \int d^3 x \lambda^{-3} (\sqrt{1+ \lambda^4(\vec B)^2}-1)
 \ee
 The condition for a finite energy static magnetic field thus takes the form
 \be
\left.\frac{dE_B(\lambda)}{d\lambda}\right|_{\lambda=1}= \int d^3 x\left [ 3-\frac{3+ (\vec B)^2}{\sqrt{1+ (\vec B)^2}}\right ]=0
 \ee
Note that whereas for a magnetic field  in Maxwell theory we have
$E(\lambda)=\int d^3 x \lambda (\vec B)^2$ and hence Derrick's
condition cannot be fulfilled, for the magnetic field in DBI  there
is no such an  obvious objection. It is easy to check that by
expanding the square root to leading order one recovers the result
in Maxwell's theory.

Next we consider the case of an electric field. For such a case we
have a scaled energy \be\label{HElambda}
 E_E= \int d^3 x \lambda^{-3}\left[ \frac{1}{ \sqrt{1- \lambda^2(\vec E)^2}}-1\right ].
 \ee
 Derrick's condition now reads
 \be
 \left.\frac{dE_E(\lambda)}{d\lambda}\right|_{\lambda=1}= \int d^3 x\left [ 3+\frac{-3+ 2(\vec E)^2}{\sqrt{(1- (\vec E)^2)^3}}\right ]=0.
 \ee
 Again, this condition does not exclude a static finite energy solution as it does  for Maxwell
 theory. Furthermore, one can easily show that the second derivative (for the
 allowed values of $0\le E^2\le 1$) is always negative, so any
 solution satisfying the above will lie at a minimum of the energy.

 For the  general case of both electric and magnetic field the scaled energy reads
 \be\label{HDBIlambda}
 E= \int d^3 x \lambda^{-3}\left[\sqrt{ 1+ \lambda^4 [\vec B)^2 + (\vec  D(\lambda))^2 + \lambda^4(\vec B\times  \vec  D(\lambda))^2] }  -1\right ],
 \ee
 where
 \be
 \vec  D(\lambda)= \frac{\lambda(\vec E + \lambda^4 (\vec E\cdot\vec B)\vec B)}{\sqrt{1+\lambda^2[\lambda^2(\vec B)^2 -(\vec E)^2 -\lambda^4(\vec B\cdot\vec E)^2]}}.
 \ee
 Upon substituting $D(\lambda)$ into (\ref{HDBIlambda}) we find that
 \be
 E= \int d^3 x \lambda^{-3}\left[ \frac{1+\lambda^4 B^2}{\sqrt{ 1- \lambda^2[(\vec E) - \lambda^2(\vec B)^2 + \lambda^4 (\vec E\cdot \vec B)^2 ]}}-1\right ],
 \ee
 which reduces to (\ref{HBlambda}) and (\ref{HElambda}) when the electric and magnetic fields are switched off, respectively.
 Derrick's condition thus takes the form
 \be
 \left.\frac{dE(\lambda)}{d\lambda}\right|_{\lambda=1}= \int d^3 x\left [ 3-\frac{3+4(\vec B)^2 -6(\vec E\cdot \vec B)^2
  -4(\vec E)^2  +(\vec B)^2((\vec B)^2 -2(\vec E\cdot \vec B)^2 ) }{\sqrt{ (1- (\vec E)^2 + (\vec B)^2 -  (\vec E\cdot \vec B)^2 )^3}}\right ].
 \ee
 Though solitons are not excluded, it is interesting to
 note that one naive ansatz, in which $\vec E$ and $\vec B$
 are orthogonal but have the same magnitude \textit{is} excluded by
 this constraint. If we take $|\vec E|=|\vec B|$ but $\vec E\cdot\vec
 B=0$, we find
 \begin{align}
\left.\frac{dE(\lambda)}{d\lambda}\right|_{\lambda=1}=-\int d^3x B^4 <0~,
 \end{align}
 so no non-trivial solutions can minimize the energy.

\section{Deformation constraints  in DBI with scalars}\label{Dbranesol}
We show more explicitly here that the Derrick or Manton constraints do not exclude the possibility of scalar solitons in the DBI action. Recall equation \ref{derrickbrane}.
 To simplify it a bit further, it will be useful to write the pullback metric as
\begin{equation}
g_{ij}=G_{ik} N^k_{\ j}=G_{ik} \left[\delta^k_{\ j} +\hat{G}^{kl}G_{ab}\partial_l Y^a\partial_j Y^b\right]=(\hat{G}N)_{ij}.
\end{equation}
The hat denotes inversion with respect to only the spatial part of the metric: $\hat{G}^{ik}G_{kj}=\delta^i_{\ j}$. Let us also define $\hat{g}^{ik}g_{kj}=\delta^i_{\ j}$. The matrix $N$ is the sum of two positive definite matrices,
\begin{equation}
 N=I+\hat{G}^{-1} (Y\cdot Y),
\end{equation}
so it is possible diagonalize it at any point in space. The
eigenvalues are of the form $1+\lambda_i$, with $\lambda_i\geq
0$.\footnote{Strictly speaking $(Y\cdot Y)$ is only semi-definite
positive ($\det(Y\cdot Y)=0$ if $\partial_i \vec{Y}$ are not
linearly independent vectors), but except in the trivial
configuration \textit{some} of its eigenvalues must be positive.}
This implies that $\sqrt{-\det g} >\sqrt{- \det G}$,  and the sum of
the first two terms in the parentheses in \eqref{derrickbrane} is
positive. In order to cancel this, the space average of the inverse
metric must satisfy
\begin{equation}
T_p\sum_l \int d^px e^{\epsilon\phi}\sqrt{-\det{g}} g^{il}(g_{il}-
G_{il}) > 0
\end{equation}
where we use the inverse of the pullback metric,
\begin{equation}
g^{\mu\nu}=\frac{1}{G_{00}-\hat{g}^{kl}G_{0k}G_{0l}}\left(
\begin{array}{cc}
1 & -\hat{g}^{jn}G_{0n} \\
-\hat{g}^{im} G_{0m} \quad & (G_{00}-\hat{g}^{kl}G_{0k}G_{0l})\hat{g}^{ij}+\hat{g}^{ik}\hat{g}^{jl}G_{0k} G_{0l}
\end{array}
\right).
\end{equation}
If we consider the case where we scale all the spatial coordinates
with the same factor $\lambda_i=\lambda$ (Derrick's condition), we
can see that this is, in principle, possible. We now have
\begin{equation}
 g^{ij}(g_{ij}-G_{ij})>0,
\end{equation}
where now summation \textit{is} implied. Then,
\begin{equation}
g^{ij} (g_{ij}-G_{ij})={\rm tr}\,(\hat{g}^{-1}(\hat{g}-\hat{G}))+\delta =p-{\rm tr}\,((N^{-1} \hat{G}^{-1})\hat{G})+\delta=p-{\rm tr}\,(N^{-1})+\delta.
\end{equation}
Where we have defined
\begin{equation}
\delta= \frac{\hat{g}^{kl}G_{0k}G_{0l}-G_{ij}\hat{g}^{ik}\hat{g}^{jl}G_{0k} G_{0l}}{G_{00}-\hat{g}^{kl}G_{0k}G_{0l}}.
\end{equation}
so the weakest condition we know must be satisfied is
\begin{equation}\label{derrickcondM}
p-{\rm tr}\,(N^{-1})+\delta >0.
\end{equation}
Since the eigenvalues of $M$ are larger than one, the eigenvalues of
$M^{-1}$ are smaller than one, so the condition \eqref{derrickcondM}
could  be satisfied by a nontrivial configuration, even in the
absence of additional fluxes. This does not, therefore, explicitly
exclude the possibility of having non-trivial scalar solitons. We reach a similar conclusion if we try to impose Derrick's condition for $G_{0i}=0$,
\begin{equation}
\sqrt{-\det g}-\sqrt{-\det G}= g^{ij}(g_{ij}-G_{ij}).
\end{equation}
This condition can be written in terms of the matrix $N$ as
\begin{equation}
\tr(N^{-1})=\frac{p}{\sqrt{\det N}}.
\end{equation}
For $p=1$ the only solution is $N=I$. For $p=2$ we already find more possibilities. In terms of the eigenvalues of $N$ we get the condition
\begin{equation}
\frac1{1+\lambda_1}+\frac1{1+\lambda_2}=\frac{2}{\sqrt{(1+\lambda_1)(1+\lambda_2)}}.
\end{equation}
This is automatically satisfied if $\lambda_1=\lambda_2$. In general for $p>1$ the conditions can be written as a polynomial equation for the eigenvalues, after squaring the original condition
\begin{equation}
\left(\sum_{i=1}^p \frac1{1+\lambda_i} \right)^2=\frac{p^2}{\prod_{i=1}^p (1+\lambda_i) }.
\end{equation}

\end{document}